\documentclass{fac}

\usepackage{url} 
\usepackage{wrapfig}
\usepackage{pdfpages}
\usepackage{bbding}
\usepackage{tikz}
\usepackage{algorithmic}
\usepackage{graphicx}
\usepackage{textcomp}
\usepackage{csquotes}
\usepackage{lineno}
\usepackage{listings}
\usepackage[normalem]{ulem}
\usepackage{framed}
\usepackage{empheq}
\usepackage{mwe}
\usepackage{caption}
\usepackage{subcaption}
\usepackage{verbatim}
\usepackage{pgfplots}
\usepgfplotslibrary{patchplots}
\usepackage{algorithm}
\usepackage{algorithmic}
\usepackage{enumitem}
\usepackage{todonotes}
\usepackage{booktabs}
\usepackage{relsize,color}
\usepackage{microtype}
\usepackage{framed}
\usepackage{multirow}
\usepackage[normalem]{ulem}
\usepackage{mdframed}
\usepackage{fancyvrb}
\usepackage{xcolor}
\usepackage{tikz}
\usepackage{algorithm}
\usepackage{algorithmic}
\usepackage{empheq}
\usepackage{float}
\usepackage{color}

\usepackage{listings}
\definecolor{dkgreen}{rgb}{0,0.6,0}
\definecolor{gray}{rgb}{0.5,0.5,0.5}
\definecolor{mauve}{rgb}{0.58,0,0.82}
\ifprodtf \else \usepackage{latexsym}\fi

\newcommand\black{\ensuremath{\blacktriangleright}}
\newcommand\white{\ensuremath{\vartriangleright}}

\newif\ifamsfontsloaded
\ifprodtf
  \newcommand\whbl{\white\kern-.1em--\kern-.1em\black}
  \newcommand\blwh{\black\kern-.1em--\kern-.1em\white}
  \newcommand\blbl{\black\kern-.1em--\kern-.1em\black}
  \newcommand\whwh{\white\kern-.1em--\kern-.1em\white}
  \amsfontsloadedtrue
\else
  \checkfont{msam10}
  \iffontfound
    \IfFileExists{amssymb.sty}
      {\usepackage{amssymb}\amsfontsloadedtrue
       \newcommand\whbl{\white\kern-.125em--\kern-.125em\black}%
       \newcommand\blwh{\black\kern-.125em--\kern-.125em\white}%
       \newcommand\blbl{\black\kern-.125em--\kern-.125em\black}%
       \newcommand\whwh{\white\kern-.125em--\kern-.125em\white}}
      {}
  \fi
\fi

\newtheorem{definition}{Definition}[section]

\title[A Game-Theoretical Self-Adaptation Framework for Securing Software-Intensive Systems]
      {A Game-Theoretical Self-Adaptation Framework for Securing Software-Intensive Systems 
      }

\author[Mingyue Zhang et al.]
    {Mingyue Zhang$^{1,2}$,
    Nianyu Li$^{1,2}$,
    Sridhar Adepu$^3$,
    Eunsuk Kang$^4$,
    Zhi Jin$^{1,2}$
    \\
    $^1$Key Lab of High Confidence Software Technologies (PKU), MoE, Peking University, Beijing, China\\
    $^2$School of Computer Science, Peking University, Beijing, China\\
    $^3$University of Bristol, United Kingdom\\
    $^4$Carnegie Mellon University, Pittsburgh, USA 
    }

\correspond{Zhi Jin, No.5 Yiheyuan Road Haidian District, Beijing, PRC. e-mail: zhijin@pku.edu.cn\\
or may also carbon copy other authors mingyuezhang@pku.edu.cn; eunsukk@andrew.cmu.edu; li\_nianyu@pku.edu.cn; sridhar.adepu@bristol.ac.uk. 
An earlier version of this paper was presented at the Fundamental Approaches to Software Engineering - 24th International Conference, FASE 2021}

\pubyear{2000}
\pagerange{\pageref{firstpage}--\pageref{lastpage}}

\begin{document}
\label{firstpage}
\makecorrespond
\maketitle

\begin{abstract}
The increasing prevalence of security attacks on software-intensive systems calls for new, effective methods for detecting and responding to these attacks. 
As one promising approach, game theory provides analytical tools for modeling the interaction between the system and the adversarial environment and designing reliable defense. 
In this paper, we propose an approach for securing software-intensive systems using a rigorous game-theoretical framework.
First, a self-adaptation framework is deployed on a component-based software intensive system, which periodically monitors the system for anomalous behaviors.
A learning-based method is proposed to detect possible on-going attacks on the system components and predict potential threats to components. 
Then, an algorithm is designed to automatically build a \emph{Bayesian game} based on the system architecture (of which some components might have been compromised) once an attack is detected, in which the system components are modeled as independent players in the game.
Finally, an optimal defensive policy is computed by solving the Bayesian game to achieve the best system utility, which amounts to minimizing the impact of the attack. 
We conduct two sets of experiments on two general benchmark tasks for security domain.
Moreover, we systematically present a case study on a real-world water treatment testbed, i.e. the Secure Water Treatment System.
Experiment results show the applicability and the effectiveness of our approach. 
\end{abstract}


\section{Introduction}
Recent progress in software-intensive systems has witnessed impressive results in many applications ranging from cloud computing system, to network infrastructure, to large-scale industrial control systems. 
These systems consist of various computational and communication devices/components that interact with each other and the environment to carry out various complex tasks.
Security attacks on software-intensive systems can result in serious consequences,
such as physical equipment damage \cite{henrie2013cyber}, large-scale blackouts \cite{cunlai2019vulnerability}, and even the loss of human life \cite{yulia2016areview}.
The increasing prevalence of attacks on these types of systems calls for more effective and systematic methods for responding to attacks and minimizing the potential harm to the environment.

We have been investigating an approach to designing the secure system based on the concept of \emph{self-adaptation}. 
Self-adaptation refers to the capability of modifying the system structure or behavior at run-time in response to changes in the environment, with the goal of continuously satisfying a desired system requirement (e.g., safety or security) \cite{DBLP:conf/dagstuhl/LemosGMSALSTVVWBBBBCDDEGGGGIKKLMMMMMNPPSSSSTWW10,salehie2009selfadaptive}. 
In the context of the secure system, the self-adaptive mechanism can be used to detect potentially on-going attacks and dynamically adjust the behavior of the system (e.g., reverse an actuator action or replace compromised components) for minimizing the impact of the attacks.  
However, there are two challenges in applying self-adaptive mechanism to security domain: (1) the adversarial nature of the environment, i.e. the attackers may carry out actions to maximize damage on the system and (2) the uncertainty of the environment, i.e. there is typically only partial information available about the attacker's actions and the parts of the system that have been compromised.

\emph{Game theory} has been used for modeling interactions between the system and the attackers as a \emph{game} between a group of \emph{players}, aiming for being able to compute an optimal \emph{policy} for the system to minimize the impact of attacks ~\cite{DBLP:journals/csur/DoTHKKBRPI17,DBLP:conf/gamesec/FarhangG16,DBLP:conf/memocode/KinneerWFGG19}. 
In particular, \emph{Bayesian games} \cite{harsanyi2004games} are designed to explicitly encode and reason about uncertainty in the data of the game (e.g., the action space, the utility, or the belief about other players).
Some of the existing works in security domain that leverage game theory~\cite{DBLP:books/daglib/0040483,DBLP:journals/csur/DoTHKKBRPI17,DBLP:conf/gamesec/FarhangG16,DBLP:conf/memocode/KinneerWFGG19} model the system as an independent player (i.e., defender). 
However, such a monolithic approach that abstracts the entire system as a single player might be insufficient for capturing certain practical scenarios, where only one part of the system is compromised while the remaining system components may cooperate with each other to mitigate the impact of an on-going attack.

We argue that compared to a coarse one-player abstraction of a system, in the context of component-based systems, modeling the defender under security attacks at the granularity of \emph{components} is more expressive. 
It allows the design of fine-grained defensive policies for the system under partial compromise.
In our prior work~\cite{nianyu2021engineering}, we propose a component-level self-adaptive framework for encoding the problem of self-adaptive security as a Bayesian game, and planning a defensive policy at the component level.
In this Bayesian game, the reward to be achieved by each player represents the desired system utility (e.g., security or performance metrics in the system), and the adversarial/abnormal players represent system components that might have been compromised by the attacker (and thus act maliciously).
Each player is assigned a type and the type of a player representing whether the player is successfully compromised or not.
The probability distribution over the types is used to encode the uncertainty of the attacks on the system.
The defensive policy for responding to attacks is generated by computing Nash equilibrium of the game.

However, there are still some limitations, hindering its application in the real-world software-intensive systems.
First of all, our prior work needs to possess a priori knowledge of the system-level utility function and the probability of each component being successfully compromised, which is difficult to obtain in practice.
Secondly, it is very laborious, time-consuming and error-prone for system developers to manually design the corresponding Bayesian game for a given system.
Moreover, the manual construction of Bayesian games cannot meet the efficiency and response time requirements of many security systems.
Thirdly, the framework proposed by our prior work is semi-automatic and when the environment changes, experts are required to participate in the design of new game models at run-time.
Many real-world software-intensive systems need to dynamically update the game models and plan new adaptation strategies. 
This requires a fully automated framework without the needs of help from the experts.

In this paper, we make an essential extension of our FASE'2021 paper \cite{nianyu2021engineering}. 
Specifically, we extend the component-level self-adaptive framework and propose automatic ways for generating the Bayesian game and computing the defensive policies. 
The main extensions and new contributions are as follows:

\begin{itemize}
    \item Our prior work presents a self-adaptive framework that incorporates Bayesian game theory to improve the resiliency of the system under potential security attacks. 
    In this framework, a system under attacks is modeled as a multi-player Bayesian game with potentially compromised players at the granularity of components, and the equilibrium of the game is used as an optimal adaptation response of the system. 
    This paper enhances the analysis process of this framework so that the system can identify the system-level utility and the compromised components through analyzing the running data at run-time rather than relying on the priori knowledge.
    Three essential improvements are incorporated into the analysis process, including the automatic calculation process of system utility, the method of assigning payoffs to each component from the system utility, and the deep learning-based predictor for compromised probability.
    
    \item We design an automatic process of generating a Bayesian game for secure systems and incorporate it into the planning process of the prior framework.
    Any system under security attacks modeled following our approach can be automatically transformed into a corresponding Bayesian game. 
    With this process, the goal of dynamically updating the game model and planning an adaptation policy can be achieved without the needs of help from human experts.
    In addition, we also propose a case-based planning method that significantly reduces the time cost.

    \item 
    To evaluate the applicability and effectiveness of the proposed approach, besides two sets of experiments on the general benchmark for security domain, i.e. a web-based client-server system called Znn.com, and an inter-domain routing system, we also conduct a comprehensive case study on an operational real-world water treatment testbed called Secure Water Treatment (SWaT) system.
    
\end{itemize}
 

The rest of this paper is organized as follows. 
Section 2 presents background and related work.
Section 3 illustrates a running example and outlines the proposed approach.
Section 4 presents the formal definition of the game for securing software-intensive systems.
Section 5 proposes the automatic process of generating the Bayesian game and computing the policy.
Section 6 conducts the experiments and discusses the experimental results.
Section 7 outlines a short conclusion and highlights the future work.

\section{Background and Related Work}
In this section, we start with background on \emph{Bayesian game theory} as well as equilibrium, and review the related work from two perspectives, i.e., securing software-intensive systems and application of game theory in security domain.

\subsection{Bayesian Game Theory}
\label{sec:back1}
\emph{Game theory} is a bag of mathematical tools designed to analyze and predict the interactions between multiple rational players. 
It assumes that these players pursue well-defined self-interests and take into account their knowledge or expectations of other players' behavior \cite{fudenberg1991game,Osborne1994A}.
A \emph{Bayesian game} is a type of game with imperfect information, that is, each player has some private information (such as action space, utility function, or belief about other players) that is only visible to itself, but instead, each of them knows the distribution over the possible information \cite{harsanyi2004games}.

\begin{definition}
A Bayesian game is a tuple $BG = \langle N, S,\mathcal{I}, A, \Theta, U, \rho \rangle$, where: 
\begin{itemize}
    \item $N$ is a finite set of players labeled by 1, 2, ..., n;
    
    \item $S$ is a finite set of states;
    
    \item $\mathcal{I}=\mathcal{I}_1\times...\times \mathcal{I}_n$ is a finite set of state partition, where $\mathcal{I}_i$ denotes the state partition of player $i$.
    $\mathcal{I}_i$ partitions state set $S$ into multiple subsets, and the states in the same subset are indistinguishable to player $i$.
    These subsets of states are named \emph{information sets};
    
    \item $A=A_1\times...\times A_n$ is a finite set of joint actions, where $A_i$ denotes a set of actions available to player $i$;
    
    \item $\Theta=\Theta_1\times...\times\Theta_n$ is a finite set of types for all players, where $\theta \in \Theta_i$ is player $i$'s type;
    
    \item $U=\{U_1,...,U_n\}$ is a set of payoff functions, where $U_i:A\times\Theta\rightarrow\mathbb{R}$ is the individual payoff function for player $i$, the individual payoff function is determined by the types of all players and actions they choose;
    
    \item $\rho:\Theta\rightarrow [0,1]$ is the joint probability distribution over types.
    $\hfill\square $
\end{itemize}
\end{definition}
Throughout the Bayesian games, we assume that the assignment of types to players is private information, while the priori type probability distribution, the action spaces and the payoff functions are assumed to be common knowledge.
The \emph{policy} for player $i$ is $\pi_i(a_i|I_i,\theta_i):\mathcal{I}_i\times\Theta_i \times A_i\rightarrow [0,1]$, and $\forall \theta_i\in \Theta_i \forall I_i\in \mathcal{I}_i,\sum_{a\in A_i}\pi_i(a|I_i,\theta_i)=1$.
A player’s policy can be pure (i.e., a policy assigning an action in $A_i$ to each information set $I_i\in\mathcal{I}_i$) or mixed (i.e., a policy giving a probability over the set of player $i$'s pure policies).
The policy is pure if it satisfies that $\forall \theta_i\in\Theta_i\forall I_i\in \mathcal{I}_i,\exists a\in A_i,\pi_i(a|I_i,\theta_i)=1$, also denoted as $\pi_i : \mathcal{I}_i\times\Theta_i \rightarrow A_i$.
The joint policy for all players is denoted as $\vec{\pi}=[\pi_1,...,\pi_n]$.\\

We now turn to the definition of \emph{Bayesian Nash equilibrium} (BNE) policy.
\begin{definition}
(Bayesian Nash equilibrium policy)
Given a joint policy for all players $\vec{\pi}^*=[\pi_1^*,...,\pi_n^*]$,  $\vec{\pi}^*$ is the Bayesian Nash equilibrium policy if and only if for any player $i\in N$, $\pi_i^*$ satisfies that:
\begin{equation}\nonumber
\setlength\abovedisplayskip{1pt}
\setlength\belowdisplayskip{1pt}
\pi_i^*=\arg\max_{\pi_i\in\Pi(\theta_i)}\sum_{\vec{\theta}_{-i}}\rho(\vec{\theta}_{-i}|\theta_i)
\mathbb{E}_{\vec{a}_{-i}\sim\vec{\pi}^*_{-i},a_i\sim \pi_i}
[U_i(a_i,\vec{a}_{-i};\theta_i,\vec{\theta}_{-i})]
\end{equation}
where $\vec{a}_{-i}=[a_1,...,a_{i-1},a_{i+1},...,a_n]$, $\vec{\theta}_{-i}=[\theta_1,...,\theta_{i-1},\theta_{i+1},...,\theta_n]$,
$\vec{\pi}_{-i}^*=[\pi_1^*,...\pi_{i-1}^*,\pi_{i+1}^*,...,\pi_n^*]$,
$\Pi(\theta_i)$ is the set of all possible policies for player $i$ under $\theta_i$,
and $\rho(\vec{\theta}_{-i}|\theta_i)$ is the conditional probability representing the player $i$'s belief about other players' types under type $\theta_i$.
$\hfill\square $
\end{definition}
Intuitively, BNE policy is a joint policy for all players, in which there is no player who can improve its profit by unilaterally modifying its policy if the actions of the rest are fixed under the constraints of uncertain players' types~\cite{fudenberg1991game,harsanyi2004games}.

\subsection{Related Work}
This paper focuses on leveraging game theory to enhance the adaptive ability of software-intensive systems for responding to security attacks.
Here, we review the available techniques and discuss applications adopting game-based mechanisms for framing our approach within the security domain.

Software-intensive systems under security attacks need to make adaptive decisions as a response to the detected threats or deviations from security goals and requirements~\cite{DBLP:conf/icse/Emami-Taba17a}.
For leading to malicious process anomalies, attackers hack into the software-intensive systems in two ways, including 1){\em Bad data injection}~\cite{gharebaghiHosseiniIzadiSafdarian,wangGuanLiuGuSunLiu}: an attacker may launch a Man-in-the-Middle (MITM) attack and send deceptive data to system components, such as Programmable Logic Controllers (PLCs).
These deceptive data may cause the system components to issue abnormal commands;
2) {\em Bad command injection}~\cite{gaoMorrisReavesRichey,meliopoulosCokkinidesFanSun,meliopoulosCokkinidesFanSunCui}: an attacker may compromise the communication links between different system components, such as the link between the controller and the controlled devices, and directly send commands to the devices. 
These attacks may lead the whole system into an unsafe state either immediately or sometime later.

Some approaches are proposed for minimizing the impact of the attack on software-intensive systems.
Bailey et al.~\cite{DBLP:conf/icse/BaileyMLYW14} generated Role Based Access Control (RBAC) models to provide assurances for adaptations against insider threats.
RBAC technique was applied to cloud computing environment to provide appropriate security services according to the security level and dynamic changes of the common resources~\cite{rolebasedcontrolincloud}.
Blount et al.~\cite{DBLP:conf/compsac/BlountTM11} designed a classifier for distinguishing the malicious data and normal data, which combines an expert system with an evolutionary algorithm, to improve the accuracy of malware detection.
Dimkov et al.~\cite{DBLP:conf/ifip1-7/DimkovPH10} discussed insider threats that span physical, cyber and social domains and present a framework to describe attacks. 

Recently, Industrial Control Systems (ICSs) has gained much attention from academia to industry. 
As a special kind of software-intensive systems, such systems are used to control industrial processes such as manufacturing, product handling, production, and distribution. 
Each ICS usually consists of sensing, driving, computing, and communication devices that interact with each other and with the environment to perform various complex tasks.
A large body of research has investigated the impact of cyber attacks on ICS's measurement and control signals~\cite{cardenas2011attacks,adepu2018assessing,liu2011false}. 
To protect the industrial control systems against cyber attacks, prior work~\cite{sabaliauskaite2017integrating,adepu2019challenges} developed various methods for monitoring, detecting and preventing cyber attacks. 
Machine learning algorithms such as support vector machines~\cite{maglaras2017novel}, $k$-nearest neighbours~\cite{al2019anomaly} and decision trees ~\cite{al2019anomaly} provide high sensitivity to attacks with precise detection.
MADICS \cite{perales2020madics}, a methodology armed with deep learning, is designed for modeling the ICS's normal/abnormal behaviors and achieving a semi-supervised anomaly detection paradigm.
Unsupervised learning methods~\cite{kravchik2018detecting,goh2017anomaly} are increasingly adopted due to their abilities to effectively detect attacks without requiring any labeled data. Li et al.~\cite{li2019mad} proposed unsupervised multivariate anomaly detection with MAD-GAN to capture the spatial-temporal correlations between sensors and actuators in the system.

However, it is notable that the application of game theory, with the characteristic of modeling the adversarial nature of security attacks and designing reliable defense with proven mathematics, has not gained the deserved attention. 
Dijk et al.~\cite{DBLP:journals/joc/DijkJOR13} proposed an imperfect-information game in which an attacker with uncertainty about its actions may periodically gain full control of an asset, with each side trying to maintain control as much as possible.
An extension work by Farhang et al.~\cite{DBLP:conf/gamesec/FarhangG16} explicitly modeled the multi-stage attacks and advanced persistent threats. 
Following this line, Kinneer et al.~\cite{DBLP:conf/memocode/KinneerWFGG19} additionally considered multiple attacker types with different goals and capabilities.
C\'{a}mara  et al.~\cite{DBLP:conf/icse/CamaraMG14,DBLP:journals/taas/CamaraMGS16} adopted a game theoretic approach and modeled the system as a turn-based stochastic game where both the system and the environment are modeled as two players.
This game models the environment as a fully competitive player aiming to minimize the system's utility and so it is a zero-sum game.
Glazier et al.~\cite{DBLP:conf/saso/GlazierG19} used a game-based approach to reason and synthesize policies for meta-manager by explicitly considering alternate potential future state for improving the performance of a collection of autonomic systems against a defined quality objective. 

Although some game-based approaches have been employed in security domain, they only act on some very general scenarios, but have not specifically conducted research on the characteristics of software-intensive systems, especially component-based systems.
In order to provide adaptive response to attacks for software-intensive systems, our prior work \cite{nianyu2021engineering} proposed a fine-grained modeling framework of the system threats and security policies at the component level. 
This paper extends this framework to realize online automatic modeling and policy calculation, thereby effectively improve its applicability and efficiency.

\section{Self-Adaptation Incorporating Bayesian Game Theory}
\label{sec:overview}
In this section, we first illustrate the main idea through a running example, i.e., the single tank system, a sub-system of the water treatment system.
Then, we give an overview of the proposed framework, i.e., \emph{self-adaptive framework incorporating Bayesian game}.

\begin{figure}
    \centering
    \includegraphics[width=4in]{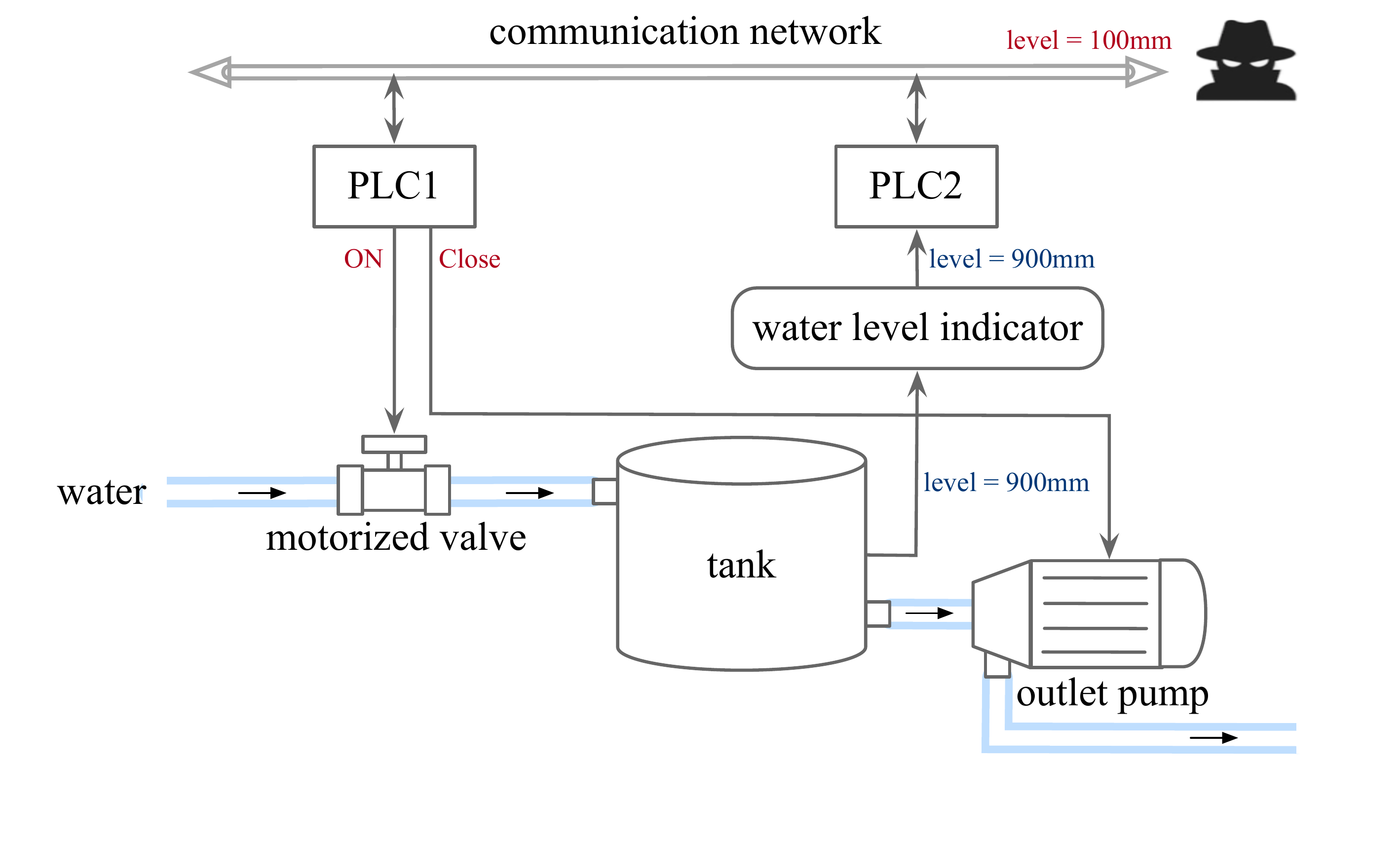}
    \vspace{-0.7cm}
    \caption{Illustrative Single Tank Example}
    \label{fig:singeltank}
\end{figure}

\subsection{Running Example}
\label{sec:runningexample}
To simplify the illustration, we adopt a simple but fully featured system with single tank (shown in Fig.\ref{fig:singeltank}). 
It is a sub-system of a real industrial control system for water treatment \cite{hau2020evaluating}.
It has two layers.
One is the physical layer consisting of a water tank, a motorized valve, an outlet pump, and a water level indicator. 
The other one is the control layer consisting of two Programmable Logic Controllers (PLC1, PLC2) which communicate with other through a wired/wireless network. 
Each PLC has its own actuators and sensors.
For example, in Fig. \ref{fig:singeltank}, PLC1 controls the valve and the pump, and PLC2 controls the indicator.
The control objective is to maintain a fixed level of water in the tank.
In the beginning, the tank is empty, and PLC1 sends command ``\text{ON}'' to valve so that the valve allows water to flow in the tank. 
After a period of time, PLC2 reads the value of the water level in the tank through the indicator, and sends the value of the water level to PLC1.
PLC1 receives the value of the water level and judges whether the level is higher than the standard level or not. 
If yes, PLC1 sends command ``\text{OFF}'' to the valve and ``\text{Open}'' to the pump so that the valve stops water flowing in the tank and the pump allows raw water to flow out the tank. 
Listing 1 shows the control policy used by PLC1 to maintain a fixed water level. 
The goal of the control policy is to maximize the utility:
\begin{equation}
\label{eq:singletank}
\centering
    Utility = -\big(
    w_1\int_{0}^{\infty} (x(t)-X_{s})^2 dt + w_2t_r + w_3(X_{max}-X_{s}) + w_4\delta
    \big)
\end{equation}
where $x(t)$ is the water level at time $t$, $X_s$ is the standard level, $X_{max}$ is the maximum water level during adjusting, $t_r$ is the time period from start time to the stable time, $\delta$ is the steady-state error (i.e., satisfying that $x(t)\in [X_s-\delta,X_s+\delta]$), and $w_1$, $w_2$, $w_3$ and $w_4$ are scaling factors. 
The first term measures cumulative deviation, the second term measures the adjusting time (or named rise time), and the third term measures overshoot.

There are some potential vulnerabilities in the single tank system.
For example, a malicious attacker attempts to damage the system, which may result in overflowing the water tank, stopping water supply, bursting the water pipes, etc.
Possible invasions generally involve the following three steps: 1) gaining access to the communication network; 2) performing reconnaissance and understanding of the process; 3) gaining control of PLCs.
After a successful invasion, the attacker can bypass the PLC and directly send the command to an actuator, or send deceptive sensor data to other PLCs.
For example, an attacker wants to overflow the water tank which follows the naive control policy shown in Listing 1. 
The attack could be \emph{bad data injection}, i.e., the attacker sends PLC1 deceptive water level data (e.g., ``level = 100mm'') which is lower than the standard level (e.g., =800mm). 
After a period of time, although the real water level in tank is 900mm ($>$800mm), due to the deceptive data (``level = 100mm''), PLC1 still keeps the motorized valve ON and the outlet pump Close, resulting in the tank overflow. 

\begin{lstlisting}[title=Listing 1: An naive Tank Water Level Control Policy, frame=shadowbox]
# Keep the water leve in (standard_level-delta,standard_level+delta)
# tank level is higher than standard level, stop inflow & start outflow
if water_level_sensor.get()>=(standard_level+delta):
    valve.Off()
    pump.Open()
# tank level is lower than standard level, start inflow & stop outflow
if water_level_sensor.get()<=(standard_level-delta):
    valve.On()
    pump.Close()
\end{lstlisting}

\subsection{Overview}
\label{sec:overview-2}
\begin{figure}
    \centering
    \includegraphics[width=5.5in]{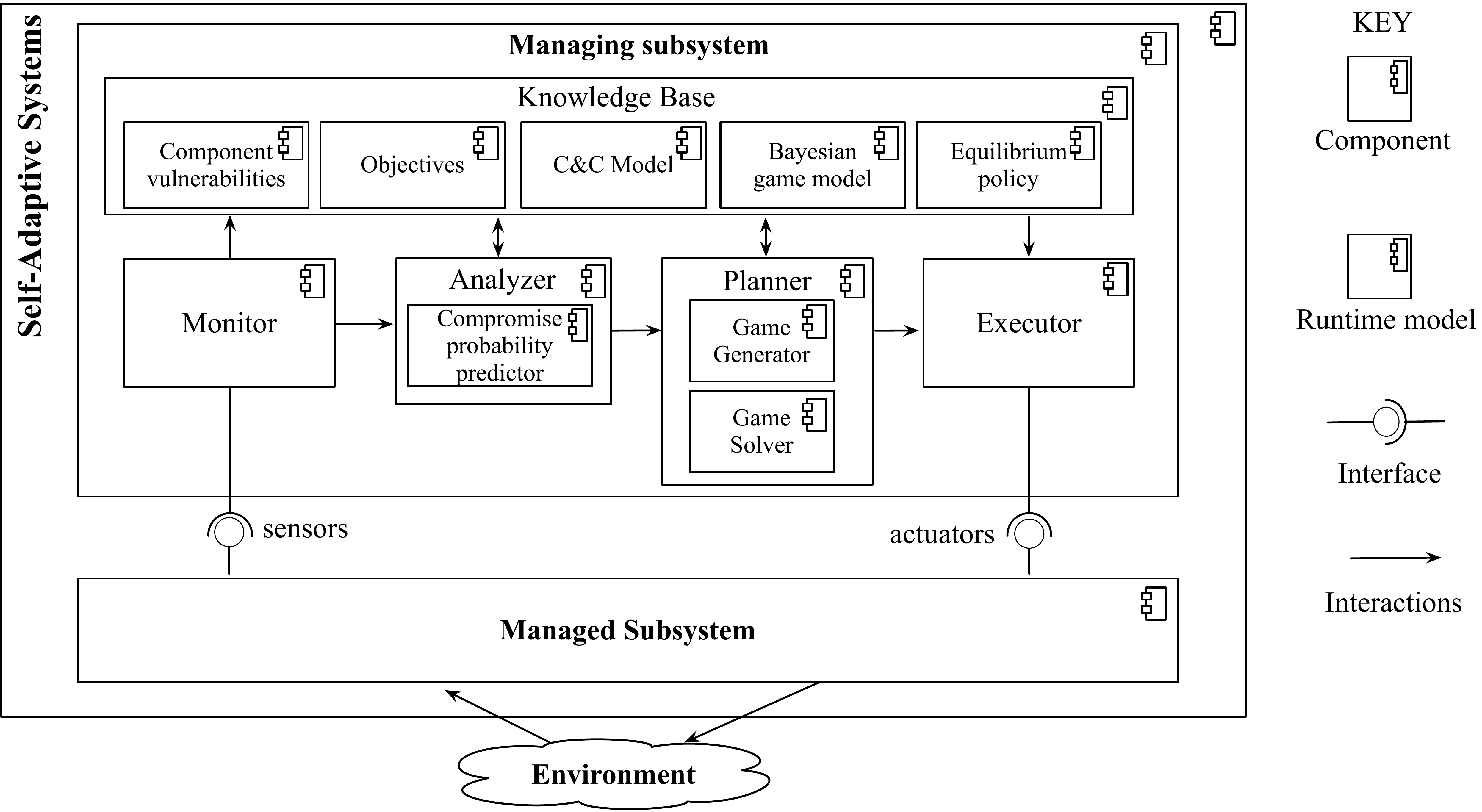}
    \caption{Self-Adaptive Framework Incorporating Bayesian Game}
    \label{fig:overviewfigure}
\end{figure}

Security attacks are usually associated with a high degree of uncertainty where the defender may know little about the identity of the attackers nor fully understand their technical effects on the system.
Bayesian games, characterized by the fact that players of a game have incomplete information about other players, are appropriate for modeling and dealing with the attacks with uncertainty.
We propose a multi-player Bayesian game to model the system under attack and to plan an optimal defensive policy for the system by solving the game. 
We further extend the typical self-adaptive framework \cite{MAPEJeffrey,DBLP:conf/icse/WeynsIS13} by incorporating the Bayesian game as shown in Fig.\ref{fig:overviewfigure} for dealing with the attacks with uncertainty.

With this framework, the system consists of two subsystems, i.e., the managed subsystem and the managing subsystem.
Back to the running example, the single tank system, which consists of sensors, actuator, other physical devices (such as water tank, communication network) and PLCs, is the managed subsystem.
It interacts with the environment, and its parameters, behaviors, or architecture can be adjusted by the command from the managing subsystem.
The managing subsystem, consisting of five parts, i.e., \emph{Monitoring}, \emph{Analysis}, \emph{Planning}, \emph{Execution}, and \emph{Knowledge Base} (MAPE-K), is designed to dynamically adjust the managed subsystem.
The key extension to the original framework \cite{nianyu2021engineering} consists of four novel parts.
They are a compromise probability predictor for predicting the probability of each component being successfully compromised, a Bayesian game model for modeling the managed subsystem under attack, a game generator for translating the system architecture description into a Bayesian game, and an equilibrium policy for responding to attacks and minimizing the potential harm. 
The four parts are organically integrated to realize the automatic conversion of the managed subsystem architecture description into a Bayesian game so that a self-adaptation policy can be generated without the intervention of humans at run-time.

With this extension, this framework can be specified as follows.

\textbf{Knowledge Base.}
To build the knowledge Base requires the system developers or domain experts to specify 
(1) the component vulnerabilities with potential behavior deviations that can be exploited by the potential attacks,
(2) the system objectives usually defined as the quality attributes quantified by the utility,
and (3) the component and connector model (i.e., C\&C model in Fig.\ref{fig:overviewfigure}) of the managed subsystem and the action space of each component.
The knowledge related to adaptation is also needed, and will be updated at run-time, including 
(4) the Bayesian game models that will be automatically generated for describing the managed subsystem under security attacks,
and (5) the equilibrium policies that will be taken as the adaptation policies for responding to attacks and minimizing the potential harm on the managed subsystem and the environment.
Knowledge Base also contains some other necessary information such as the history of system behaviors and environment information.

\textbf{Monitor.}
Monitor receives events generated in the managed subsystem or environment. 
These events indicate the execution of system actions or natural changes in the environmental factors. 
Monitor also gathers and synthesizes the on-going attacks information through sensors and conveys the information to the Knowledge Base. 
For example, events such as loss of large number of user request or command injection may indicate potential attacks on the web server.

\textbf{Analyzer.}
Speculative analysis is for identifying the conditions of the environment/managed subsystem that indicate the violations or or better fulfillment of goals, which may arise based on the information from Monitor.
The Analyzer performs analysis and further estimates the probability that certain components are attacked; potential deviated malicious actions are identified; and the rewards for the attack are estimated, based on the knowledge about component vulnerabilities and system objectives. 
Such attack probabilities can be obtained with a statistical combination of all feasible scenarios along with expert judgment~\cite{DBLP:conf/eisic/DudorovSN13,probabilityanalysis}.
Moreover, in the analyzer, a deep learning-based predictor is designed to predict the probability of being compromised (i.e., compromised probability predictor in Fig.\ref{fig:overviewfigure}).
This predictor is offline trained on the data set of the system running history, and takes as input the system running data from monitor and outputs compromised probabilities of the components at run-time.

\textbf{Planner.}
Planner generates the workflow of adaptation actions aiming to counteract the violations of system goals or to better achieve the goals. 
The adaptation actions are generated by automatically solving the Bayesian game that is derived from (1) the compromised probability; (2) the C\&C model; along with (3) the system objectives. 
The changes in the three factors are reflected in the Bayesian game model at run-time, and ultimately lead to the changes in adaptation actions.
In order to realize the automatic generation and solving of the Bayesian game, we design a game generator and a game solver, which will be elaborated in Section~\ref{sec:AutomaticProcess}.
An adaptation equilibrium will be generated if a security threat exists to deal with the unexpected attacks, or if the violation cannot be handled, it prompts to make proper changes to the system. 
For example, distributing more user requests to normal servers while reducing user requests to those with a high probability of compromise, as well as adjusting the fidelity for servers could be feasible actions for website-based client-server system under security attacks~\cite{DBLP:conf/icse/ChengGS09a,DoSJavier}.

\textbf{Executor.}
During execution, the policies from the adaptation equilibrium are enacted on the managed subsystem through the actuators in the Executor. 
Typical examples in the single tank system (as shown in Fig.\ref{fig:singeltank}) could be changing the state of the motorized valve or the outlet pump.

The Monitor and the Executor follow the design principle of the MAPE-K loop and we assume adequate monitoring and execution environment in place through which the selected adaptation policies are enacted. 
Thereupon, in the next two sections, we will focus on planning activity with Bayesian game theory, and the analysis methods on potential attacks with uncertainties.

\section{Bayesian Games for Component-Based Systems Under Attacks}
This section first defines the component-based systems under attacks.
Second, the multi-player Bayesian game is formally figured out.
Finally, we describe the system-level utility, and propose a \emph{Shapley Value}-based approach to automatically deduce the component's payoff from the system-level utility\footnote{Here, the utility is at the system level, and the payoff is at the component level.}.

\subsection{Component-Based System Under Attacks}

Many software-intensive systems are complex and large-scale and  \emph{component-based approach} is commonly adopted for designing such systems~\cite{heineman2001component}.
Within a component-based system, a component is an artifact that is one independent and replaceable part of which a more composite component is made up and any part can be separated from or attached to the system. 
Moreover, any component fulfills a clear function in the context of a well-defined architecture.
In our running example in Fig.~\ref{fig:singeltank}, typical examples are the \emph{water Pump}(P), the \emph{tank water Level Indicator/Transmitter}(LIT), the \emph{Motorized Valve}(MV), and the \emph{water Tank}(T).
Components forming architectural structures affect different quality attributes.
For example, the quality attributes of cumulative deviation, adjusting time, and overshoot identified in the single tank example are influenced by the actions of the three components (i.e., the motorized valve, the outlet pump, and the water level indicator) and characterized as a function as shown in Eq.(\ref{eq:singletank}) in Section \ref{sec:runningexample} mapping them to utility values.

\begin{definition}
\label{def:ics}
A component-based system can be defined as a tuple $S = \langle C, A, E, Q\rangle$, where:
\begin{itemize}
    \item $C$ is a finite set of $n$ components labeled by $1,2,...,n$;
    \item $A$ is a set of joint actions $A=A_1\times...\times A_n$, where $A_i$ denotes a action set available to component $i$;
    \item $E \subseteq C \times C$ is a finite set of directed edges which defines the connections, i.e. message flow relations, between components.
    Edge $\langle i,j\rangle$ denotes that component $i$ connects to component $j$;
    \item $Q$ is a set of quality attributes that a system is interested in; for each $Q_x\in Q$, there is a subset of components $SubC_x \subseteq C$ each of which contributes to $Q_x$.
    $\hfill\square $
\end{itemize}
\end{definition}


Normally, each component is trying to make the right reaction to maximize the system utility, essentially like a rational player in the game theory. 
Naturally, a system under normal operation could be viewed as a cooperative game dealing with how coalitions interact. 
Each component can be modeled as an independent player and the interacting components/players form a coalition.
For instance, in the running example, three components, i.e., the motorized valve, the outlet pump, and the water level indicator, cooperate with each other to maintain the safe water level in the tank and to continuously pump out water.
When the outlet pump crash due to an attack, other components need to take defensive actions to avoid tank overflowing.

The security attacks on the component-based system ultimately act on some of the components.
Accordingly, instead of modeling an attacker or several attackers with possible complex behaviors over different parts of the system, we model the on-going attacks the system is enduring at the component level since the vulnerabilities of the components as well as their potential behavior deviations are comparatively easy to observe.
The attack model, a model for formally describing how the attacker attacks the system and the possible effects of the attack, will be obtained by synthesizing the information from Monitor and Analyzer, and the algorithm for generating the model will be elaborated in Section \ref{sec:att}.
When a system is attacked, we call it a component-based system under attacks.

\begin{definition}
\label{def:att}
Given a component-based system $S = \langle C, A, E, Q\rangle$, the component-based system under attack of $S$ is $S_{att} = \langle C_a, C_n, A, A_{att}, E, Q\rangle$, where:
\begin{itemize}
    \item $C_a\subseteq C$ is a set of abnormal components affected by the attack, $C_n\subseteq C$ is a set of normal components, $C_a\cap C_n=\emptyset$, and $C_a\cup C_n=C$;
    
    \item For each $a_i\in C_a$, $A_{att,a_i}$ is the set of $a_i$'s actions controlled by attack, i.e. attack changes the actions of $a_i$, and we say that $a_i$ takes competitive actions. 
    For each $n_i\in C_n$, its action set is still $A_{n_i}$, i.e. attack does not change the actions of $n_i$, and we say that $n_i$ takes cooperative actions;
    
    \item $A_{att}=A_{att,a_1}\times\cdots\times A_{att,a_l}\times A_{n_1}\times\cdots\times A_{n_m}$ is the set of joint actions of the component-based system under attack, in which $l=|C_a|$, $m=|C_n|$, $a_1, \cdots, a_l\in C_a$ and $n_1, \cdots, n_m\in C_n$.
\end{itemize}
\end{definition}


According to the definition, when a system is under an attack, the components affected by the attack may behave abnormally, but the others behave normally.
Table \ref{attakctype} lists some typical attacks on the single tank system and the possible effects of these attacks.
For example, in Table \ref{attakctype}, attack A2 is carried out in the following way: if the ``indicator'' detects the real water level is higher than the standard level, the attacker uses the ``indicator'' to send out LOW signal; otherwise, to send out HIGH signal.
The motorized ``valve'' and ``outlet pump'' need to take right action (ON/OFF, Open/Close) based on not only the the water level signal (LOW or HIGH) but also the probability of the ``indicator'' being successfully attacked.
In this case, $C_a=\{indicator\}$, $C_n=\{valve, pump\}$, $A_{indicator}=\{HIGH,LOW\}$, $A_{valve}=\{\text{ON},\text{OFF}\}$, $A_{pump}=\{\text{Open},\text{Close}\}$,
$A_{att,indicator}$ represents the abnormal behavior of attacked component ``indicator'', i.e., $A_{att,indicator}=\{H-L,L-H\}$, where $H-L$ ($L-H$) means sending LOW signal (HIGH signal) when the real water level is higher (lower) than the standard level.

\begin{table}
\centering
\small
\caption{Typical Attacks on Component-based System.}
  \begin{tabular}{ccccl}
    \toprule
    Attack ID & Attack description & Affected components ($C_a$) & Possible effects\\
    \midrule
    A1 & Keep the outlet pump always Close & outlet pump & tank overflow \\
    A2 & Falsify the status of the water level & water level indicator & tank underflow or overflow \\
     A3 & Control the PLC1 & motorized valve, outlet pump & stopping water supply\\
    \bottomrule
  \end{tabular}
   \label{attakctype}
\end{table}

\subsection{Definition of Bayesian Games}
According to Definition \ref{def:att}, a corresponding multi-player Bayesian game for the component-based system under attack is designed as follows.
Each (normal or abnormal) component is considered as one of the players in a Bayesian game.
In order to be consistent with the description of the component-based system, in the Bayesian game we use the term ``component'' to refer to the player that the component acts as.
To encode the uncertainty of the attack (e.g., the attack may or may not be successful, or from the perspective of effects, the component may or may not be compromised), we model it as the probability distribution over components' types.
In the Bayesian game, each component owns one or two types.
For a normal component, its type is deterministically set to \emph{cooperative}.
For an abnormal component, its type is stochastically set to \emph{cooperative} or \emph{competitive}.
If a component is cooperative, it will cooperate with other cooperative components to maximize the system-level utility.
If a component is competitive, it will independently maximize its own payoff.
The following is defining such a Bayesian game.

\begin{definition}
\label{def:sas}
Given a component-based system under attack $S_{att} = \langle C_a, C_n, A, A_{att}, E, Q\rangle$, the multi-player Bayesian game is defined as a tuple $G = \langle N, \Theta, A, A_{coo}, A_{com},U, P, T,\mathcal{I} \rangle$, where:
\begin{itemize}
     \item $N=\{\text{Nature}\}\cup C_a\cup C_n$, in which \emph{Nature} is a fictitious player, and the others are the components in $S_{att}$.
     The elements in $N$ are labeled by 0, 1, 2, ..., n, where $0$ denotes \emph{Nature}, and 1, 2,..., n denotes those components;

    
     \item $\Theta=\{\theta_{coo},\theta_{com}\}$ is a set of types for the components, where $\theta_{coo}$ represents a cooperative type and $\theta_{com}$ represents a competitive type;
    
    \item $A_{coo}=A_{coo,1}\times...\times A_{coo,n}$ is a finite set of joint actions, where $A_{coo,i}$ denotes a finite set of cooperative actions of component $i$;
    
    \item $A_{com} = A_{com,1}\times...\times A_{com,n}$ is a finite set of joint actions, where $A_{com,i}$ denotes a finite set of competitive actions of component $i$;
    
   \item $U=\{U_1,...,U_n\}$ is a set of component-level payoff functions, where $U_i:\Theta\times A \rightarrow\mathbb{R}$ is a payoff function for component $i$;
   
     \item $P=\{p_i|i\in C_a\cup C_n\}$ is a set of probabilities, where $p_i$ is the probability of component $i$ being type $\theta_{com}$, and $1-p_i$ is the probability of component $i$ being type $\theta_{coo}$;
    
    \item $T$ is a rooted game tree, and the node of the game tree is also called state.
    The tree satisfies that: (1) at each non-terminal state of the tree, only one player is allowed to take an action; 
    (2) each component only takes action at most once on all paths from the root state to the terminal state;
    
    \item $\mathcal{I}=\mathcal{I}_1\times...\times \mathcal{I}_n$ is a finite set of state partitions, where $\mathcal{I}_i$ denotes the state partition of component $i$.
    $\mathcal{I}_i$ partitions the set consisting of all states where component $i$ takes action in tree $T$ into multiple subsets, and the states in the same subset are indistinguishable to component $i$.
    $\hfill\square $
\end{itemize}
\end{definition}

The Bayesian game will be generated from the models of component-based systems under attack.
In the following sections, we will detail the generation of the Bayesian game.
Concretely, given $S_{att} = \langle C_a, C_n, A, A_{att}, E, Q\rangle$, the corresponding Bayesian game will be obtained step by step as follows:
\begin{itemize}
    \item For each $c_i\in C$, the corresponding player's cooperative action set $A_{coo,i}$ is $A_i$; for each $n_i\in C_{n}$, the corresponding player's competitive action set $A_{com,i}$ is $A_i$; for each $a_i\in C_{a}$, the corresponding player's competitive action set $A_{com,i}$ is $A_{att,a_i}$;
    
     \item The system-level utility function, denoted as $U_0$, is designed based on the quality attributes $Q$ (such as Eq.(\ref{eq:singletank}) in Section \ref{sec:runningexample}). 
     We assume that $U_0$ satisfies a condition that: $\forall \vec{a}\in A, U_0(\vec{a}) \equiv \sum_{i\in C_n}U_i(\theta_{coo},\vec{a})$.
     Section \ref{sec:component-payoff} will further detail how to automatically compute each component's payoff function $U_i$ from the system-level utility.
     
     \item The probability set $P$ is generated by the compromise probability predictor which will be elaborated in Section \ref{sec:att};
    
    \item Game tree $T$ determines the order in which the players take action in the Bayesian game, and it is generated from $C$, $A$, $E$, and $P$ through a generation process which will be further elaborated in Section \ref{sec:generating};
    
    \item State partition set $\mathcal{I}$ depends on $C_a$ and $P$, and the process of generating $\mathcal{I}$ will be further elaborated in Section \ref{sec:generating}.
\end{itemize}

For instance, in the running example (shown in Fig.\ref{fig:singeltank}), three physical components are explicitly modeled as players, i.e., motorized valve, outlet pump, and water level indicator. 
They are denoted by $c_1$, $c_2$, and $c_3$, respectively.
The action sets of them are $A_1=\{\text{ON},\text{OFF}\}$, $A_2=\{\text{Open},\text{Close}\}$, and $A_3=\{\text{LOW},\text{HIGH}\}$. For $c_3$, the action LOW/HIGH means that the water level indicator sends LOW/HIGH signal to PLCs.
Assume $c_2$ is under attack, i.e., $C_n=$\{1, 3\}, $C_a$=\{2\} and the attack causes $c_2$ to only take action \text{Close}, i.e., $A_{att,2}=$\{\text{Close}\}.
The connection between components is given as $E=\{\langle 3,1\rangle, \langle 1,2\rangle\}$.
In the corresponding Bayesian game, $A_{coo,1}=\{\text{ON},\text{OFF}\}$, $A_{coo,2}=\{\text{Open},\text{Close}\}$, $A_{coo,3}=\{\text{LOW},\text{HIGH}\}$;
$A_{com,1}=\{\text{ON},\text{OFF}\}$, $A_{com,2}=\{\text{Close}\}$, $A_{com,3}=\{\text{LOW},\text{HIGH}\}$.
We assume that the probability of $c_2$ being competitive is $0.8$.
Based on $C$, $A$, $E$, and $P$, the game tree $T$ is generated as shown in Fig.\ref{fig:attack1}:
first, the indicator detects the water level and takes action; second, the valve takes action; third, the nature stochastically determines whether the attack on the pump is successful or not, i.e., choosing the type of the pump (competitive or cooperative); fourth, the pump takes action, if the pump is cooperative, it takes action based on its payoff; otherwise, it takes action Close.
For $c_3$, $\mathcal{I}_3$ partition its states into one set, i.e., $1:1$ in Fig.\ref{fig:attack1}.
For $c_1$, $\mathcal{I}_1$ partition its states into two sets, i.e., $2:1$ and $2:2$ in Fig.\ref{fig:attack1}.
For $c_2$, $\mathcal{I}_2$ partition its states into eight sets, i.e., $3:1$, $3:2$,..., $3:8$ in Fig.\ref{fig:attack1}.
In Fig.\ref{fig:attack1}, each terminal state of the game tree has an n-tuple of payoffs, meaning the component-level payoff functions, i.e., $U_i$.



\subsection{Modeling Component-level Payoff}
\label{sec:component-payoff}

\textbf{System-level Utility.}
Given a component-based system $S = \langle C, A,E,Q\rangle$, the system-level utility is a function $F: dom(Q)\rightarrow\mathbb{R}$ that assigns a real number (the utility value) to each possible combination of the quality attributes, where $dom(Q)$ denotes the range of $Q$\footnote{In this paper, the utility is based on the von Neumann–Morgenstern (VNM) utility theorem
\cite{morgenstern1953theory}. VNM proved that ``any individual whose preferences satisfied the four axioms (i.e., completeness, transitivity, continuity, independence) has a utility function; such an individual's preferences can be represented on an interval scale'' (refer to \url{https://en.wikipedia.org/wiki/Von_Neumann-Morgenstern_utility_theorem}).}.
For example, Eq.(\ref{eq:singletank}) assigns a utility value to the combination of the cumulative deviation, the adjusting time and the overshoot.
For any system, the values of all quality attributes depend on the joint action of all components, i.e. the system-level utility function is defined as $U_0:A\rightarrow\mathbb{R}$.

However, due to different roles of the components and the complex relationship between them, it is intractable to manually design an appropriate component-level payoff function.
Since the normal components try to maximize the system-level utility and the abnormal components minimize the system-level utility or maximize their own payoffs, the payoff functions for normal and abnormal components are modeled in different forms.

\textbf{Normal Component's Payoff.}
In order to automatically deduce the payoff of each normal component from the system-level utility, we adopt \emph{Shapley Value Method}. 
It involves fairly distributing both gains and costs to several players working in coalition proportional to their marginal contributions \cite{Osborne1994A,shapley1953vaule}.
Given a component set $C$, and a \emph{feature function} $v$, the Shapley value of normal component $i$ is:
\begin{equation}
\setlength\abovedisplayskip{1pt}
\setlength\belowdisplayskip{1pt}
\label{eq:shapley}
    \phi(C,v,i)=\frac{1}{|C|!}\sum_{C'\subseteq C\verb|\|\{i\}}|C'|!(|C|-|C'|-1)![v(C'\cup\{i\})-v(C')]
\end{equation}
where $|C|$ is the cardinal number of set $C$; $C\verb|\|\{i\}$ is the set $C$ excluding component $i$, feature function $v$ associates nonempty subset $C'$ of $C$ with real number $v(C')$.
$v(C')$ values the expected system-level utility when the system only controls the components in $C'$.
Our previous work arbitrarily takes an average utility function as $v(C')$ \cite{nianyu2021engineering}. 
In this paper, we give a more appropriate method to calculate $v(C')$ from $U_0$:
\begin{enumerate}
    \item For any component $c_i\in -C'$ (here, $-C'=\complement_C C^{'}=\{c|c\in C \wedge c\notin C'\}$ is the complementary set of $C^{'}$), its policy is set to random, i.e., $\forall a\in A_i, \pi_i(a)=\frac{1}{|A_i|}$;
    
    \item The feature function $v(C')$ is calculated with: 
    \begin{equation}
    \label{eq:featurefunction}
        \setlength\abovedisplayskip{1pt}
        \setlength\belowdisplayskip{1pt}
        v(C')=\arg\max_{\vec{a}_{C'}\in A^{C'}}\mathbb{E}
    \Big[
    U_0([\vec{a}_{-C'},\vec{a}_{C'}])\Big|\vec{a}_{-C'}\sim \pi_{-C'}
    \Big]\
    \end{equation}
    where $U_0(\vec{a})$ is the system-level utility function evaluating the system utility when joint action $\vec{a}$ is taken, $A^{C'}$ is the joint action space of $C'$, $\vec{a}_{-C'}$ and $\vec{a}_{C'}$ are the joint actions taken by components in $-C^{'}$ and $C^{'}$, respectively, $\pi_{-C'}(\vec{a}_{-C'})=\prod_{c_i\in -C^{'}}\pi_i(a)$ is the joint policy for components in $-C^{'}$.
\end{enumerate}

The design of $U_0(\vec{a})$ depends on the application scenario. 
For a simple application, $U_0(\vec{a})$ is normally manually designed.
For a complex application, $U_0(\vec{a})$ is implemented by the following three steps.
First, a simulator is designed to simulate the relation between control signals/actions and the quality attributes.
Second, the values of the quality attributes (e.g., throughput and water/chemical dosing level) are obtained by running the simulator.
Third, the payoff function is designed by domain experts to map the quality attributes to utility value, such as Eq.(\ref{eq:singletank}).
On the basis of Shapley value, the payoff function for normal component $i$ is:
\begin{equation}
\label{eq:normal}
  U_i(\theta_{coo},\vec{a})=U_i(\theta_{com},\vec{a})=U_0(\vec{a})\frac{\phi(C,v,i)}{\sum_{j\in C_n}\phi(C,v,j)}  
\end{equation}

\textbf{Abnormal Component's Payoff.}
The abnormal component is a kind of component which may be compromised.
In this paper, if the abnormal component's payoff is given as Definition \ref{def:att}, we will directly use it; otherwise, we will consider fully competitive scenarios, and model the system as a zero-sum game.
For the latter, if abnormal component $i$ is successfully compromised (i.e., its type is competitive), we assume that it is trying to minimize the system-level utility, and hence its payoff function is modeled as:
\begin{equation}
    U_i(\theta_{com},\vec{a}) = -\frac{1}{|C_a|} U_0(\vec{a})
\end{equation}
If abnormal component $i$ is not compromised (i.e., its type is cooperative), that means that it is trying to cooperate with other normal components.
Hence, its payoff function is modeled as: 
\begin{equation}
\label{eq:coo}
\begin{split}
    U_i(\theta_{coo},\vec{a})=&\mathbb{E}_{j\in C_n}[U_j(\theta_{coo},\vec{a})]=
    \mathbb{E}_{j\in C_n}
    [
    U_0(\vec{a})\frac{\phi(C,v,j)}{\sum_{k\in C_n}\phi(C,v,k)}
    ]\\
    =& U_0(\vec{a}) \mathbb{E}_{j\in C_n}
    [\frac{\phi(C,v,j)}{\sum_{k\in C_n}\phi(C,v,k)}
    ]
    = \frac{1}{|C_n|}U_0(\vec{a})
\end{split}
\end{equation}

Notice that: (1)``Abnormal component'' refers to a component being under attack; (2) The attack may succeed or fail. 
If the attack succeeds, the component is competitive or compromised; otherwise, the component is cooperative; (3) If the component is competitive, it maximizes $U_i(\theta_{com},\cdot)$; otherwise, it maximizes $U_i(\theta_{coo},\cdot)$; (4) the classical definition of payoff function for component $i$ is $U_i:\Theta\times A$, and since in this paper, the types of other components do not change the payoff for component $i$,
$U_i$ can be defined as $U_i:\Theta_i\times A$ for notation simplicity.

The following is an example of system utility allocation for the single tank example (shown in Fig.\ref{fig:singeltank}).
$C$ is $\{\text{motorized valve},\text{outlet pump},\text{water level indicator}\}$, which are labeled by 1, 2, and 3, respectively.
The action sets are $A_1=\{\text{ON},\text{OFF}\}$, $A_2=\{\text{Open},\text{Close}\}$, and $A_3=\{\text{LOW},\text{HIGH}\}$.
The compromised probabilities of the three components are $p_1=0,p_2=0,p_3=0.8$, hence $C_n=\{1,2\}$, and $C_a=\{3\}$.
For calculating $v(\{1,3\})$, the following three steps need to be conducted:

\begin{enumerate}
    \item Corresponding to Eq.(\ref{eq:featurefunction}), $C'=\{1,3\}$ and $-C'=\{2\}$. Component 2's policy (i.e., $\pi_{-C'}$ in Eq.(\ref{eq:featurefunction})) is set to random, i.e., $\pi_{2}(\text{Open})=\pi_{2}(\text{Close})=\frac{1}{2}$;
    
    \item $A^{C'}$ in Eq.(\ref{eq:featurefunction}) is $A_1\times A_3=\{\text{ON},\text{OFF}\}\times \{\text{LOW},\text{HIGH}\}$.
    All the possible joint actions $\vec{a}\in A_1\times A_3$ are taken in the simulator, and the simulator simulates the corresponding quality attributes. Eq.(\ref{eq:singletank}) maps the quality attributes to a real number, which is $U_0(\vec{a})$.
    
    \item $v(\{1,3\})=\arg\max_{a\in {\{\text{ON},\text{OFF}\}}\times \{\text{LOW},\text{HIGH}\}}(\frac{1}{2}U_0([a,\text{OFF}])+\frac{1}{2}U_0(a,\text{ON})$.
\end{enumerate}

For clarity, some notations used above are summarized in Table \ref{tab:notation}.

\begin{table}
\centering
  \caption{Some Important Notations}
  \label{tab:notation}
  \begin{tabular}{ccl}
    \toprule
    Notation & Meaning\\
    \midrule
    $C$ & the component set \\
    $C_a$ & the abnormal component set \\
    $C_n$ & the normal component set \\
    $E \subseteq C \times C$ & the set of directed edges between two components\\
    $A=A_1\times ....\times A_n$ & the joint action set, and $A_i$ denotes the set of actions available to component $i$\\
    $A_{att}$ & joint action set of the component-based system under attack\\
    $A_{coo}$ & joint action set of the cooperative components\\
    $A_{com}$ & joint action set of the competitive components\\
    $P = \{p_i|i\in C\}$ & the probability set, and $p_i$ is the probability of component $i$ being compromised \\
    $U=\{U_1,...,U_n\}$ & the component-level payoff function set, $U_i$ denotes the payoff function for component $i$\\
    $U_0$ & the system-level utility function\\
    $v:2^C\rightarrow \mathbb{R}$ & the feature function of a set of components, $2^C$ denotes the power set of $C$\\
    $\phi(C,v,i)$ & the Shapley value function\\
    $-C'=\{c|c\in C \wedge c\notin C'\}$ & the complementary set\\
    $\Theta=\{\theta_{coo},\theta_{com}\}$ & the type set of components, $\theta_{coo}$ is cooperative type, and $\theta_{com}$ is competitive type\\
    $\pi_i:\mathcal{I}_i \times A_i\times \Theta \rightarrow [0,1]$ & the policy of component $i$\\
    \bottomrule
  \end{tabular}
\end{table}

\section{Generation Process of Bayesian Game}
\label{sec:AutomaticProcess}
In this section, we start by the attack model which describes the objective/intention and capabilities/resources of an attacker.
Based on the attack model, a deep learning-based classifier is built to predict the compromise probability of each component.
And we design a process to automatically transform the component-based system under attack modeled using our approach into a Bayesian game.
Finally, we analyze the time complexity of the automatic game generation as well as the game solving processes, and propose a workaround to reduce the time spent at run-time.

\subsection{Predicting Potential Attacks}
\label{sec:att}
To predict the probability of each component being successfully compromised, we design a deep learning-based classifier, which classifies the components into normal and abnormal, as well as gives the probability of the abnormal component being compromised (i.e., the probability of competitive type $p_i$).

First, we introduce an \emph{attack model} for the component-based system under attack.
Malicious attackers are able to exploit vulnerabilities to enter the system and access communication channels and computing devices \cite{rocchetto2016attacker}.
We assume only a proper subset of components can be manipulated by the malicious attackers.

\begin{definition}
\label{def:threat}
Given a component-based system $S=\langle C,A,E,Q\rangle$, the \emph{attack model} is formally defined as a tuple $AT=\langle Obj, Cap,A_{att},\pi_{att}\rangle$, where:
\begin{itemize}
    \item $Obj$ is a set of attacker's objectives;
    \item $Cap\subset C$ is the attacker's capability, and it is a proper subset of components in the component-based system;
    \item $\pi_{att}:Obj\times Cap \rightarrow A_{att}$, is the attack policy of an attacker, where $A_{att}$ is attack action space (refer to Definition \ref{def:att}).
    $\hfill\square $
\end{itemize}
\end{definition}
The attacker needs to take action $a\in A_{att}$, which affects the status of the components of the system.
We assume that it is possible to detect the potential attack by analyzing the dynamics of these statuses.
One solution to the detection of attacks is as follows: (1) designing a \emph{process invariant}, which is a mathematical relationship among the run-time status (e.g., water level, the rate of flow, and concentration of chemical dosing) of the system; (2) monitoring the run-time status to check if the process invariant is violated.
If the status violates the process invariant, we can infer that some parts of the system are under attack.
For example, in our running example, one of the process invariant is the water flow, i.e., when the tank water level is constant, the water flowing in is equal to the water flowing out. 
Once the system detects a violation of the process invariant, it will reason out the compromised components.
In \cite{sridhar2016using,sridhar2017water}, the design of process invariant is derived manually and requires a lot of domain knowledge, hindering the application in practice.

For automatic detection of attacks, a logical analysis of data (LAD) method is introduced in \cite{das2020anomaly}.
Under normal circumstances, i.e., when the system is operating in a steady state, the status of sensors follows a typical pattern (such as the water flow invariant).
The LAD method is used to automatically generate rules from these sensor statuses. 
With these rules, LAD-based analyzer can identify potential deviations and localize the anomaly.
LAD classifier is essentially a special binary classifier, which divides the components into two categories, i.e., normal and abnormal. 
There are two phases: 
In the first phase, prime patterns, i.e., positive (the normal state of the system) and negative patterns (the anomalous state of the system), are extracted from historical observations data. 
Then, the two patterns are used to build two rule-based classifiers, which can be used to classify either ``under attack'' or ``semi-normal'' state. 
In the second phase, the rule-based classifiers are used to detect abnormal activity, and calculate the probability of a component being compromised.

In this paper, on the basis of the attack model and the LAD classifier, we further propose a deep learning-based classifier to automatically identify the compromised components and predict the attack probability of these components.
Formally, the classifier is defined as $CLF: S_C \rightarrow [0,1]^{|C|}$, where $S_C$ is the status space of all sensors, $C$ is the component set, $[0,1]^{|C|}$ denotes the joint compromised probability space
$\underbrace{[0,1]\times...\times[0,1]}_{|C|}$.
The classifier is implemented with Algorithm \ref{CLF}.
It takes as input the sensor status $s_c\in S_C$ and outputs the compromised probability list $p$ and an indicator list $t$. 
$p[i]$ denotes the probability of component $i$ being successfully compromised, $t[i]=0$ or $t[i]=1$ denotes that component $i$ is normal or abnormal.
In Algorithm \ref{CLF}, line 3 gets the ID of a given component.
In line 4, $Clf_{i}(\cdot;\theta)$ is a deep neural network, where $\theta$ is the weights of the network, and it is used to generate the compromised probability of component $i$. 
For each component $i$, there is a corresponding deep neural network $Clf_{i}(\cdot;\theta)$.
The loss function of $Clf_{i}(\cdot;\theta)$ is mean-square error between the classifier's output and the ground truth. 
The network structure of $Clf_{i}(\cdot;\theta)$ is designed as follows:
(1) the input layer is a fully-connected layer with size [20,30], and the activation function is Rectified Linear Unit (ReLU);
(2) there is a hidden layer, which is a fully-connected layer with size [30,10], and the activation function is ReLU; and
(3) the output layer is a Softmax layer with size [10,2].
The training data set for each $Clf_{i}(\cdot;\theta)$ is constructed as follows:
(1) a piece of training data consists of $s_c$ and a label of component $i$ (normal or not abnormal); and
(2) positive examples are generated through collecting the running data when the SWaT system runs normally, and negative examples are generated through simulating the attack on component $i$ in the SWaT simulator and collecting the data in the simulator. 
In this paper, we collect 5,000 pieces of data by using the SWaT simulator designed by Hau et al. \cite{hau2020evaluating} as the training data set.

$Clf_i(;\theta)$ is trained in the offline phase, as the LAD-based analyzer does.
In Line 5 of Algorithm \ref{CLF}, $Out$ is the output of $Clf_i(;\theta)$, and it is a  two-dimension tensor. $Out[0]$ and $Out[1]$ are the probabilities of a component being compromised and not being compromised, respectively. 
The output layer of $Clf_i(;\theta)$ is Softmax, which ensures that $Out[0]+Out[1]\equiv 1$.
$TH \in [0,1] $ in Line 6 is a hyperparameter as a threshold, and its value is negatively correlated with the attacker's capability $Cap$.
Intuitively, if $Cap \equiv C$, $TH$ should be set to $0$; if $Cap \equiv \emptyset$, $TH$ should be set to $1$.

\begin{algorithm}[htb]
\algsetup{linenosize=\small}
\caption{CLF($s_c$)}
\label{CLF}  
\begin{algorithmic}[1]
\STATE Initialize a compromised probability list $p$, and initialize a type list $t$;
\FOR {$c\in C$}
\STATE $i\leftarrow \text{GetID(\emph{c})}$;
\STATE Feed $s_c$ to $Clf_i(\cdot;\theta)$;
\STATE Get $Out[\ ]$ from $Clf_i(s_c;\theta)$;
\IF {$Out[0]\leq TH$}
\STATE $t[i] \Leftarrow 0$, $p[i] \Leftarrow 0$;
\ELSE
\STATE $t[i] \Leftarrow 1$, $p[i] \Leftarrow Out[0]$;
\ENDIF
\ENDFOR
\STATE Return $p,\ t$;
\end{algorithmic}
\end{algorithm}


\subsection{Generating Bayesian Game and Computing Equilibrium Policy}
\label{sec:generating}
In order to automatically generate a Bayesian game for a component-based system under attack and compute the defensive policy by solving the game, we design a process in the planning activity.
Given a component-based system under attacks $S_{att} = \langle C_a, C_n, A, A_{att}, E, Q\rangle$, the corresponding \emph{Bayesian game}
is converted by the following five steps  \footnote{Our project link is: \url{https://github.com/GeorgeDUT/GameForSAS}}:
\begin{enumerate}
    \item \textbf{Generate the modeled component graph:}
    Based on system architecture $\langle C,E\rangle$, three actions are conducted: (1) those components without run-time actions are removed, and any pair of replaceable components are combined; (2) the two components which are connected through removed components are directly connected by an edge; and (3) the components in $C_a$ are labeled in the refined system architecture graph denoted as $\langle C',E' \rangle$.
    
    \item \textbf{Generate the order of play:} A TOPOLOGICAL-SORT algorithm \cite{cormen2009introduction} takes as input an architecture graph $\langle C',E' \rangle$, and outputs the order of play for all components, denoted as $L=[c_i,c_j,...,c_k]$, where $c_x\in C'$, and that $c_i$ is in front of $c_j$ means $c_i$ takes action before $c_j$.
    For each component $c_i$ in $L$, if $c_i\in C_a$, then a \emph{nature} player (denoted as $\mathbb{N}$) will be inserted in front of $c_i$. Notice that nature is a virtual player without payoff, and it is used to determine the type of a component.
    
    \item \textbf{Generate the component-level payoff:} 
    The payoff for each normal component is allocated with system-level utility function by the Shapley Value Method ($\phi(C',v,i)$).
    The payoff for each abnormal component being compromised and being not compromised are set to $U_i(\theta_{com},\cdot)$ and $U_i(\theta_{coo},\cdot)$, respectively.  
    
    \item \textbf{Establish the Bayesian game and generate the Gambit file:} An ADD-NODE algorithm (Algorithm \ref{algorithm_add}) takes as input a root node and the length of $L$, and recursively calls itself to generate a game tree. 
    In Algorithm \ref{algorithm_add}, $c.player \in C'$ denotes the ID of the component, $p(c.player)$ denotes the probability of component $c.player$ being compromised, i.e., $p_{c.player}$, and $|L[d].A|$ is the the cardinal number of component $L[d].player$' action space.
    A DLR\footnote{DLR is a kind of tree traversal algorithm. D, L, and R mean that visiting root node, left node, and right node, respectively.} algorithm (Algorithm \ref{algorithm_dlr}) takes as input a root node of the game tree generated by ADD-NODE algorithm, and outputs a list of all nodes in the tree. 
    Based on the list of nodes, an XML formation file for Gambit is automatically generated.
    In Algorithm \ref{algorithm_dlr}, $waiting[-1]$ in line 5 means the last element in queue $waiting$; the function MAX in line 13 represents finding a child node with maximal ID.
    
    \item \textbf{Solve the Bayesian game}: The XML game file constructed is put into a game solver, \emph{Gambit} \cite{gambit}, to find a pure Nash equilibrium, which, in essence, is the safest reaction for the system to potential attacks.
    Gambit is an open-source collection of tools for building both strategic game and extensive game models, computing Nash equilibrium and analyzing game results.
\end{enumerate}

\begin{figure}
    \centering
    \includegraphics[width=6in]{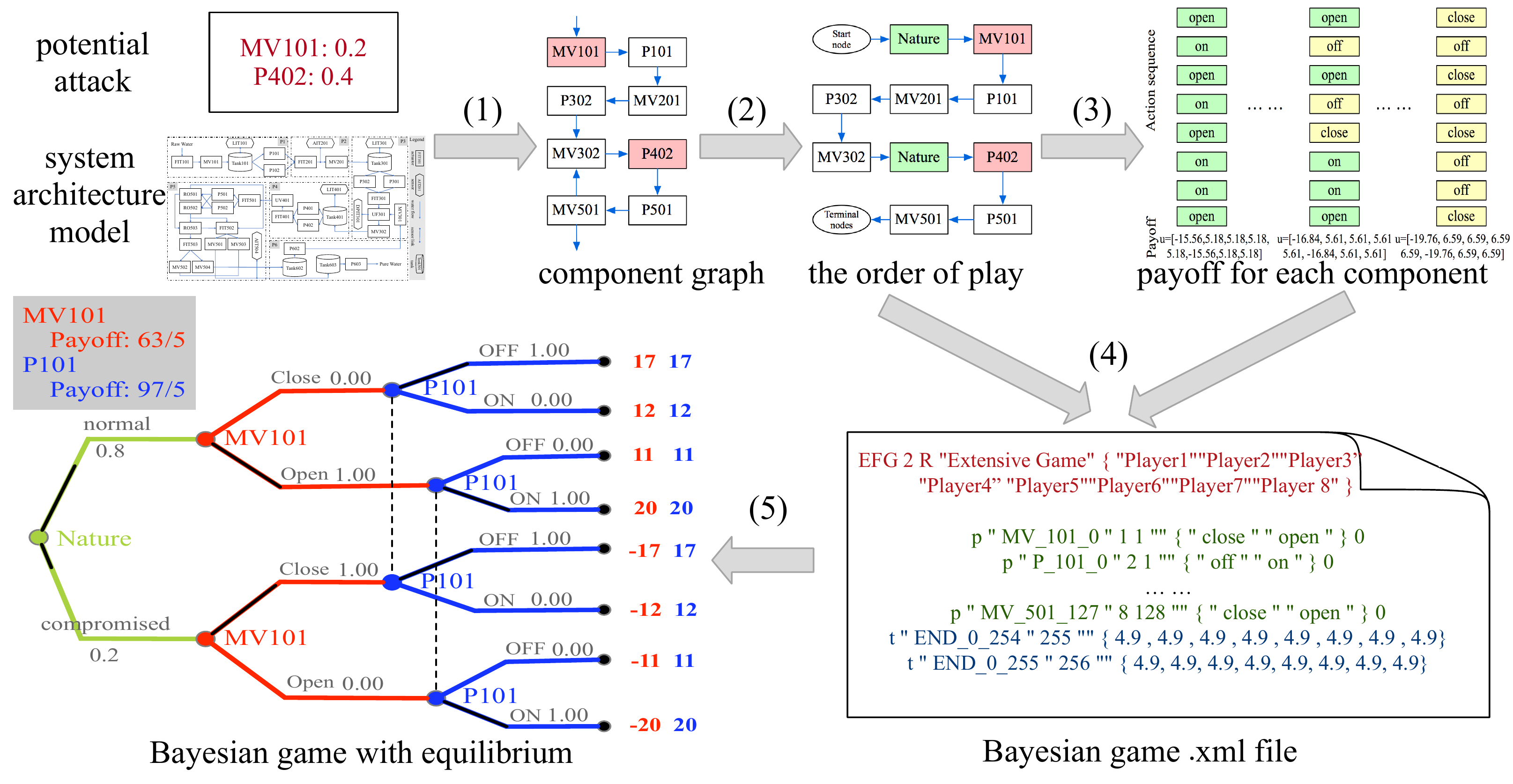}
    \caption{The Planning Activity: Automatic Process of Game Translation}
    \label{fig:construct}
\end{figure}

\begin{algorithm}[htb]
\algsetup{linenosize=\small}
\caption{ADD-NODE($c$,$d$)}
\label{algorithm_add}  
\begin{algorithmic}[1]
\IF {$d>=|L|-1$}
\FOR{$i=0$ \textbf{to} $\Big| L[d].A \Big|-1$ }
\STATE Set \emph{node} to terminal node, and set its payoff;
\STATE \emph{c.child}[\emph{i}] $\Leftarrow node$;
\ENDFOR
\ELSE
\FOR{$i=0$ \textbf{to} $\Big|L[d].A\Big|-1$ }
\STATE Set \emph{node} to non-terminal node, and set \emph{node.player} $\Leftarrow$ L[$d+1$].$player$;
\IF {\emph{c.player} is $\mathbb{N}$}
\STATE Set \emph{c.probability}=[p(\emph{c.player}),1-p(\emph{c.player})];
\ENDIF
\STATE \emph{c.child}[$i$] $\Leftarrow node$;
\STATE ADD-NODE(\emph{c.child}[$i$],d+1);
\ENDFOR
\ENDIF
\end{algorithmic}
\end{algorithm}

\begin{algorithm}[htb]
\algsetup{linenosize=\small}
\caption{DLR(\emph{root})}
\label{algorithm_dlr}  
\label{algorithm2}  
\begin{algorithmic}[1]
\STATE Initialize queue \emph{waiting} $\Leftarrow \emptyset$ and \emph{NodesList} $\Leftarrow \emptyset$;
\STATE \emph{start} $\Leftarrow$ \emph{root};
\STATE \emph{waiting}.push(\emph{start});
\WHILE{$waiting \neq \emptyset$}
\STATE $start \Leftarrow waiting[-1]$;
\STATE $NextNodesList \Leftarrow start.children$;
\FORALL{$node \in start.children$}
\IF{$node$ is visited}
\STATE Remove $node$ from $NextNodesList$;
\ENDIF
\ENDFOR
\IF{$NextNodesList \neq \emptyset$}
\STATE $m \Leftarrow$ MAX$(start.children)$;
\STATE $NodesList.push(m)$;
\STATE $waiting.push(m)$;
\STATE Set $m$ to visited;
\ELSE
\STATE waiting.pop();
\ENDIF
\ENDWHILE
\STATE Return \emph{NodesList};
\end{algorithmic} 
\end{algorithm}

The time complexity of the automatic generation process is analyzed as follows:
\begin{itemize}
    \item Step 1 takes time $\Theta(|C|)$ to remove and combine the components, takes time $\Theta(|E|)$ to set new edges, and takes time $\Theta(|C|)$ to label the potentially compromised (i.e., abnormal) components.
    
    \item Step 2 calls the TOPOLOGICAL-SORT algorithm, and the time complexity is  $\Theta(|C|+|E|)$.
    
    \item Step 3 needs to calculate the Shapley value through Eq.(\ref{eq:shapley}).
    Eq.(\ref{eq:shapley}) takes a constant time (which depends on the application) to evaluate the system-level utility, takes time $\Theta(|A_i|^C)$ to compute the feature function $v(C')$, and takes time $\Theta(|C|!)$ to compute the Shapley value for each component.
    Since the joint action $\vec{a}$ does not affect Shapley value $\phi(C',v,i)$, we can calculate the Shapley value once and reuse it in Eq.(\ref{eq:normal}).
    The time complexity of Step 3 is $\Theta(|A_i|^C|C|!)$.
    
    \item Step 4 calls the ADD-NODE algorithm recursively many times to establish an N-ary tree, where $N=\max_{i\in C}{|A_i|}$. 
    The running time $T(n)$ of ADD-NODE algorithm can be described by the recurrence $T(n)=NT(n/N)+O(1)$, where $n$ is the number of nodes in the tree. 
    Based on master method \cite{cormen2009introduction}, $T(n)=\Theta(n)$. 
    Because the N-ary tree contains all possible action sequences, the nodes of the tree is $num = \prod_{i=1}^{|C|}|A_i|$, and $\min_{i\in C}|A_i|^C \leq num \leq \max_{i\in C}|A_i|^C$. The total time spent for step 4 is $\Theta(N^C)$.
    
    \item Step 5 performs a game solver to compute the Nash equilibrium, which is the most time-consuming of all steps.
    Briefly speaking, the complexity of computing Nash equilibrium is PPAD (polynomial parity argument, directed version). 
    For two-player games, the problem of finding a Nash equilibrium is a linear complementarity problem, and it can be solved by the Lemke-Howson algorithm \cite{lemke1964equilibrium}.
    For n-player games, a simplicial subdivision method has been proposed for computing the Nash equilibrium \cite{mckelvey1996computation}. 
    In this paper, to efficiently and automatically figure out the equilibrium policy as the adaptation response, we adopt Gambit as the game solver.
    The time spent in solving the game by using Gambit is approximately exponential in the number of all components.
\end{itemize}

The automatic generation process has prohibitive complexity, hindering the application in online real-time planning.
Fortunately, an effective workaround is to perform the generation process in the offline phase and store the Nash equilibrium into the knowledge base.
A piece of knowledge case is defined as follows:

\begin{definition}
\label{def:kcase}
A piece of knowledge case is a tuple $kc=\langle \vec{p}_{com},\vec{a}^* \rangle$, where:
\begin{itemize}
    \item $\vec{p}_{com} = [p_1,...,p_n]$ is a probability vector, where $n=|C|$, and $p_i$ is the probability of component $i$ being successfully compromised;
    \item $\vec{a}^* =[a_1,...,a_n] \in A$ is the equilibrium joint action derived by the equilibrium policy, where $a_i$ is the equilibrium action taken by component $i$.
\end{itemize}
\end{definition}

In the offline phase, we set different compromised probabilities $\vec{p}_{com}$, compute the corresponding equilibrium policy, and store the equilibrium results as knowledge cases in the knowledge base denoted as $KB$.
In the online phase, the analyzer predicts the compromised probability $\vec{p}_{com}'=[p_1',...,p_{n}']$, and the planer retrieves the most similar case $kc$ in the knowledge base, i.e., $kc = \arg\min_{kc\in KB}\sum_{j=1}^{|C|}(kc(p_j)-p_j')^2$, where $kc(p_j)$ denotes the compromised probability of component $j$ in case $kc$.
Then, the equilibrium joint action of $kc$ will be taken as the defensive action and carried out by the executor.
The time spent in the case-based planning depends on the efficiency of retrieval.
And by using B-trees, AVL trees, or other storage-retrieve techniques, the time spent in the planning can be greatly reduced to $O(\text{lg}n)$.

\section{Evaluation}
To validate the proposed approach, we conduct experiments in three cases: (1) a web-based client-server system; (2) an interdomain routing system; and (3) a real water treatment system.
First, we assess the effectiveness of our approach in the web-based client-server system.
Our approach is used to enhance the ability to maintain load balancing even when some servers are under attack.
Second, we extend our approach to a system with more complex network topology, i.e., the interdomain routing system, model the system under attack as a routing game, and design a dynamic programming algorithm to solve the game.
Third, to evaluate the applicability of our approach, we systematically conduct a case study on a real-world water treatment testbed called Secure Water Treatment (SWaT) system.
It turns out that the proposed approach can achieve the self-adaptation goal and outperform other baseline approaches.

\subsection{Analysis Results for Znn.com}
\label{sec:e1}
\subsubsection{Znn.com System}
Znn.com is a hypothetical news website that has been used as a representative system for the application of self-adaptive systems~\cite{DBLP:conf/icse/ChengGS09a,DoSJavier}. In a typical workflow, given a request from a client, the web server fetches appropriate content (in form of text) from its back-end database and generates a web page containing a visualization of the text. Furthermore, the system also provides an optional service with multimedia content (e.g., images, videos). This service involves additional computation on the server side, but also brings in more revenue compared to the requests with only text. 
With $R_M$ and $R_T$ being the revenue, $C_M$ and $C_T$ being the computation of one response to a user request with the media content and with only text content, respectively, we assume that $R_M > R_T > 0$ and $C_M > C_T > 0$.

\begin{figure}
    \centering
    \includegraphics[width=2.2in]{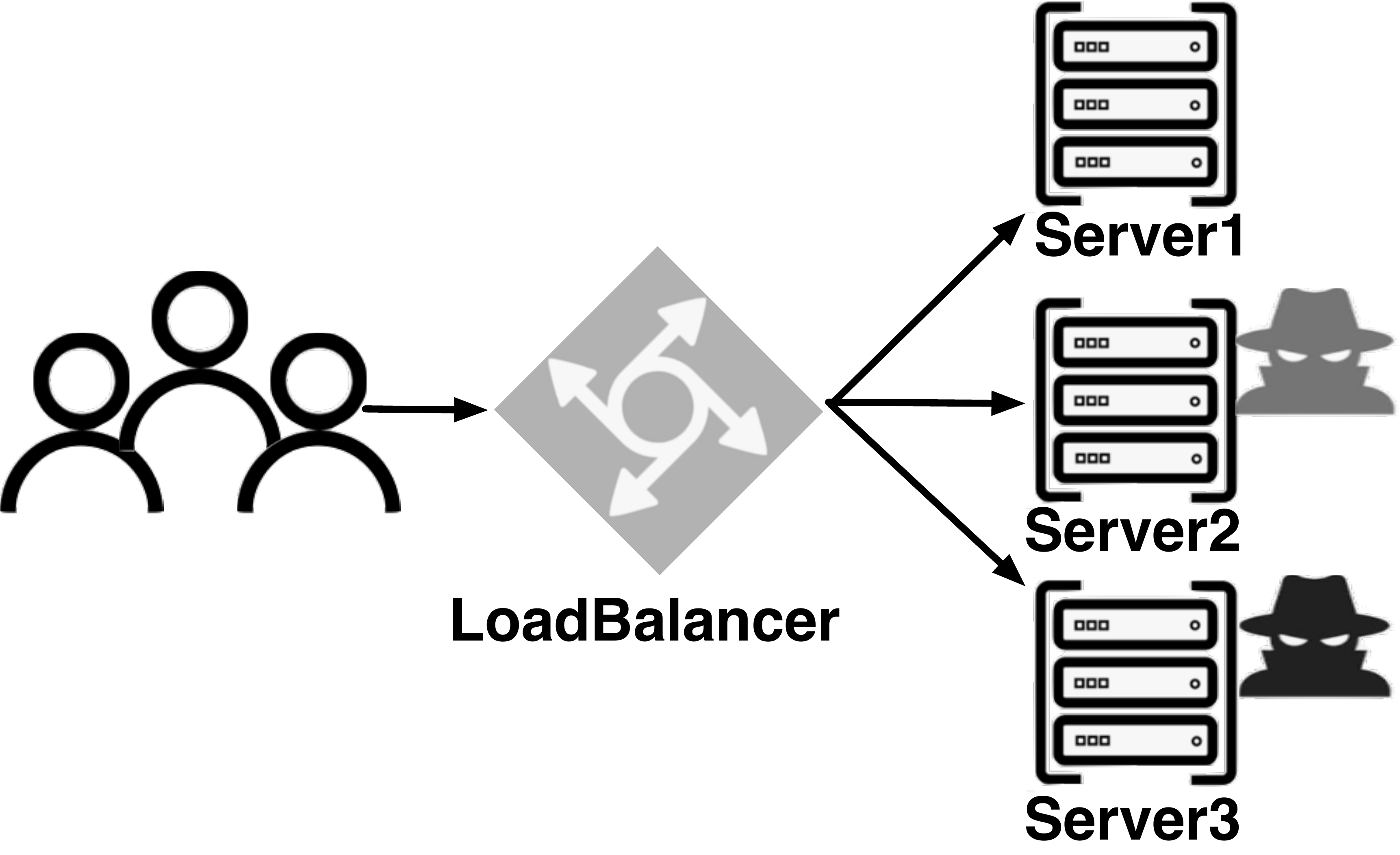}
    \caption{An example of Znn.com system}
    \label{runningexample}
\end{figure}

In order to support multiple servers, a \emph{LoadBalancer} is added to distribute the requests from the users to a pool of servers, as shown in Fig.~\ref{runningexample}. The cost of each server is proportional to its load due to, such as potential high response time since companies such as Amazon, eBay, and Google claim that increased user perceived response time results in revenue loss~\cite{DBLP:conf/nsdi/LloydFKA13}. To be more specific, the cost per server is denoted by $(S_i - T)^2/K$ where $S_i$ is the current occupied load for server $i$, depending on the request serving mode (i.e., $S_i = D_iC_T$ in text only while $S_i = D_iC_M$ in multi-media mode where $D_i$ is the number of requests distributed to server $i$); $T$ is the threshold beyond which the response time would be affected; $K$ is a constant used to adjust the cost ratio.  

The goal of the self-adaptive system is to maximize the difference between revenue and cost. 
\begin{equation}
\label{eq:ut-server}
\setlength\abovedisplayskip{0pt}
\setlength\belowdisplayskip{1pt}
    U = R_Mx_M + R_Tx_T - \sum\limits_{i=1}^3 (S_i\leq T\ ?\  0\ :\ (S_i - T)^2/K)
\end{equation}
where $x_M$ and $x_T$ are the numbers of responses with media and text content, respectively; the penalty is the sum of the cost for all three servers. 

Suppose that some of the servers are vulnerable to various attacks such as password guessing, SQL injection, command injection, etc\footnote{The following website details the web server and the types of attacks: \url{https://www.greycampus.com/opencampus/ethical-hacking/web-server-and-its-types-of-attacks}.}. 
The information collected from the web server, however, cannot fully demonstrate its compromise due to, e.g., the deficiencies of scanning tools, but with uncertainty.
As shown in the figure, \emph{Server2} could be potentially attacked with a 20\% probability while \emph{Server3} is with a higher probability of 50\%. These two servers, if compromised in reality, might perform harmful actions controlled by the attackers to achieve their objectives, rendering the loss of system reward. 
Here we assume the malicious policies of simply discarding all the distributed user requests. The reward of attacks is denoted by the system loss, i.e., subtracting the maximum reward the system could achieve from the reward under attacks, leading to a zero-sum game. 

\subsubsection{Experiment Setup and Results}
\begin{figure}
    \centering
    \includegraphics[width=5.3in]{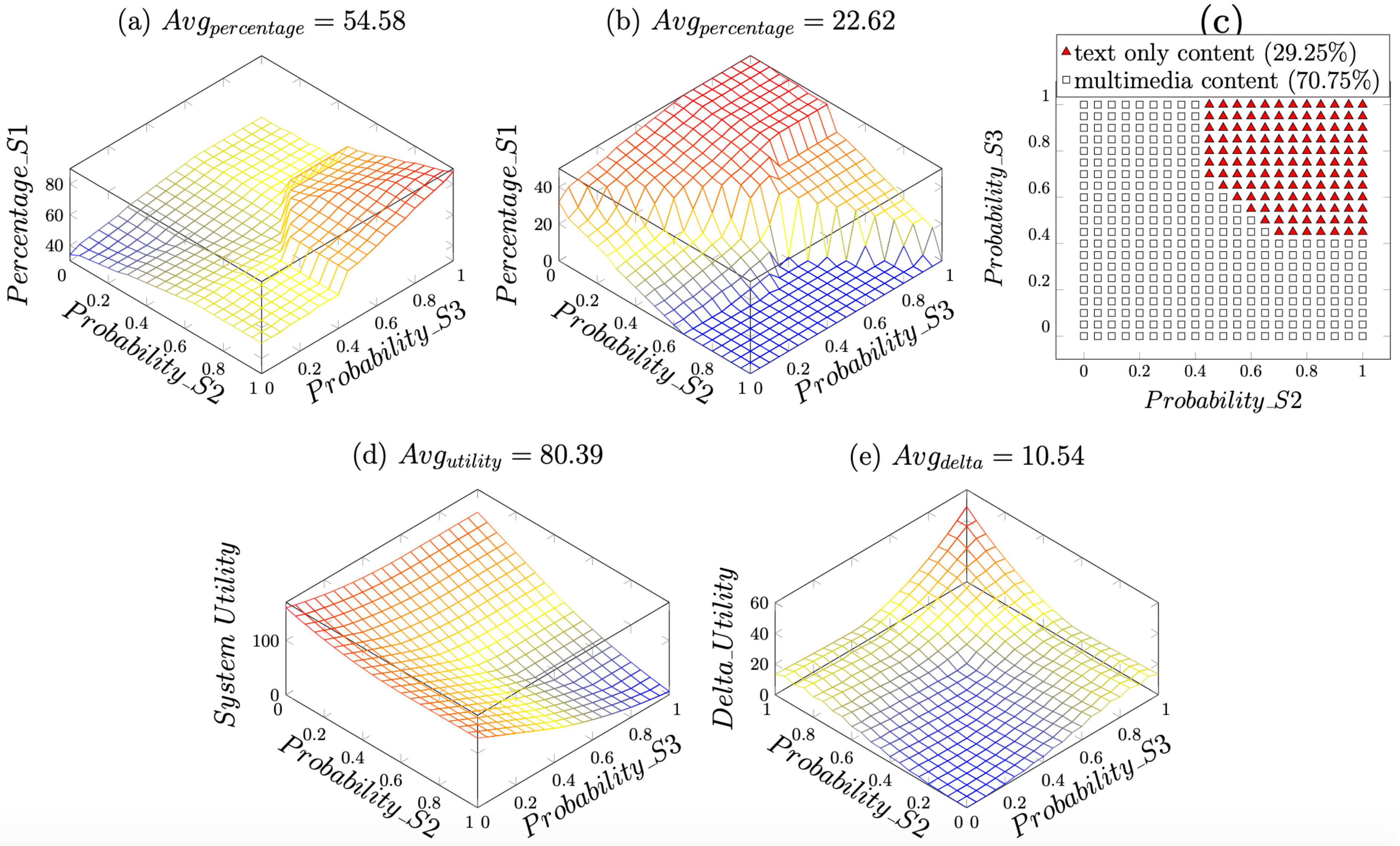}
    \caption{Results for Znn.com website: (a) percentage of user requests to \emph{Server1};  (b) percentage of user requests to \emph{Server2}; (c) policies for \emph{Server1}; (d) system utility with game theory approach; (e) delta utility between Bayesian game theory approach and probabilistic model checking approach.}
    \label{fig:runningresult}
\end{figure}

In this part, we demonstrate how our approach can produce adaptation decisions under security attacks for Znn.com website to enhance the system utility. 
In particular, we exploit the Bayesian game by following the aforementioned steps and generate the equilibrium. 
To explore different attack scenarios, we statically analyze a discretized region of the state space, which is projected over two dimensions that vary the malicious probability (i.e., $probability\_S2$ and $probability\_S3$) of \emph{Server2} and \emph{Server3} respectively (with values in the range [0, 1]). 
Each state of the discrete set requires a solution of the game with the Nash equilibrium that quantifies the best utility the system could obtain. 
The experiment takes less than one minute to generate all the results, as shown in Fig.~\ref{fig:runningresult}, and for each state, the solution generation time is negligible. 
To set up the experiment, we assume there are 100 user requests - the maximum load of a server in text only mode - with $R_M=1.6$, $R_T=1$, $x_M = 1.4$, $x_T = 1$, $T=50$, and $X=25$ in Eq.(\ref{eq:ut-server}). 
Additionally, we adopt the probabilistic model checking method as the baseline \cite{DoSJavier,DBLP:conf/icse/CamaraMG15,nianyulExplanation} and compare our Bayesian game theory method with it in terms of the system utility. 

Fig.~\ref{fig:runningresult} (a) illustrates the percentage of user requests distributed to \emph{Server1} from the policy for the \emph{LoadBalancer} in equilibrium. 
As expected, the percentage of \emph{Server1} increases progressively with the increasing malicious probability of \emph{Server2} and \emph{Server3} as more user requests are supposed to be processed by a server under normal operation.
In particular, we observe that the user percentage is around one third when both \emph{Server2} and \emph{Server3} are functioning normally (i.e., both $probability\_S2$ and $probability\_S3$ are 0), with \emph{LoadBalancer} equally delivering the user requests since none of the servers is compromised.
Moreover, the percentage for \emph{Server1} reaches around 84\% when the other two servers are fully compromised. 
In this situation, \emph{LoadBalancer} does not deliver all user requests to \emph{Server1}; otherwise \emph{Server1} may be overloaded with the  increasing costs due to high response time  which in turn outweigh its benefits of request processing. 

Fig.~\ref{fig:runningresult} (b) describes the percentage of user request that \emph{LoadBalancer} delivers to \emph{Server2} in the equilibrium. 
We can also observe that user requests to \emph{Server2} are negatively proportional to its malicious probability. 
Particularly, user requests are 50 when probability $probability\_S2$ is 0 while \emph{Server3} is fully malicious (i.e., $probability\_S3$=1) where \emph{LoadBalancer} should equally distribute the user request to both \emph{Server1} and \emph{Server2}.
Fig.~\ref{fig:runningresult} (c) presents the policy in equilibrium for \emph{Server1}. 
The states in which text content is provided are indicated by red triangles, whereas the multimedia policies for \emph{Server1} are denoted by white rectangles. 
As we can see, red points are in the upper right corner where malicious probabilities of \emph{Server2} and \emph{Server3} are greater than 50\%, which means that they are very likely compromised. 
Therefore, \emph{LoadBalancer} distributes as many user requests as possible to \emph{Server1}, thus \emph{Server1} choosing to provide text only content in avoid of overloading. 
Otherwise, \emph{Server1} can provide multimedia content in less load condition to promote user satisfaction with higher revenue.  

Fig.~\ref{fig:runningresult} (d) illustrates the maximum utility the system can achieve under various attack situations.
In particular, we observe that the utility reaches around 160 when all three servers are cooperative and is progressively decreased with the increasing malicious probability of \emph{Server2} and \emph{Server3}. 
This is consistent with the fact that the system utility is deteriorated under security attack. 
To compare the system utility in game theory with existing methods, we adopt probabilistic model checking~\cite{probabilisticModelChecking} as the comparison standard to formally model the running example and synthesize the adaptation policy maximizing its expectation of the utility by reasoning about reward-based properties ~\cite{DoSJavier,DBLP:conf/icse/CamaraMG15,nianyulExplanation}. 
Fig.~\ref{fig:runningresult} (e) presents the delta between two approaches (i.e., system utility with game theory approach minus the utility with the probabilistic model checking approach). Without security attacks, the adaptation decisions generated by the two approaches achieve the same utility. 
However, with the increasing malicious probability of \emph{Server2} and \emph{Server3}, game theory approach outperforms, providing the better response to make up for the utility loss due to security attack, and the average delta is 10.54, i.e., 15\% outperforming with the average utility 80.39 achieved by game theory.

\subsection{Analysis Results for Routing Games}
\label{sec:e2}
To evaluate our approach and assess its applicability for validation, we consider a set of experiments on an interdomain routing application. 
We first define the game (Section \ref{sec:routinggame}) and propose a dynamic programming algorithm to solve the equilibrium by decomposing the problem into smaller and tractable sub-games (Appendix \ref{sec:dp}). 
The results are present (Section \ref{sec:eval2}) with a sensitivity analysis, illustrating how the system can choose a robust policy effective for a range of threat landscapes, and a utility analysis by quantifying the defender's utility with Bayesian game compared to a greedy solution within the security context. 

\begin{figure}
\centering 
\includegraphics[width=5.1in]{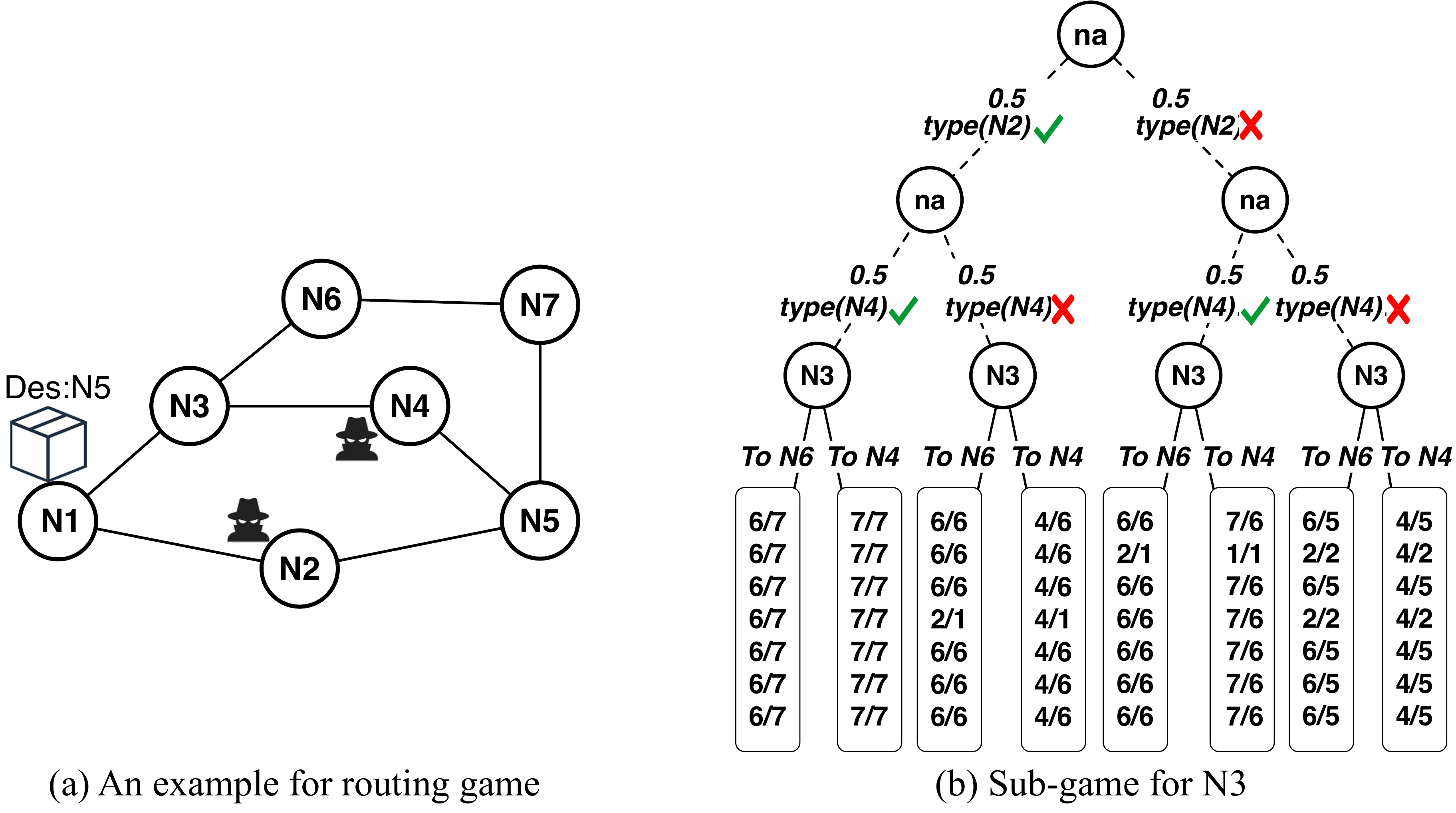} 
\caption{Interdomain Routing System}
\label{routingexample}
\end{figure}

A routing system is usually composed of smaller networks called nodes as shown in Fig.~\ref{routingexample}(a). 
Since not all nodes are directly connected, packets often have to traverse several nodes and the task of ensuring connectivity between nodes is called interdomain routing~\cite{routing1,routing2}.
Each node could be owned by economic entities (Microsoft, AT\&T, etc.) and might be compromised by the attacker at any time. 
Therefore, it is natural to consider interdomain routing from a game-theoretic point of view. 
Specifically, game players are source nodes located on a network, aiming to send a package (i.e., starting at $N1$) to a unique destination node (i.e., $N5$). 
The interaction between players is dynamic and complex - asynchronous, sequential, and based on partial information - and the best policy for each player as the adaptation response is updated as needed.

\subsubsection{Game Definition for Interdomain Routing System}
\label{sec:routinggame}
The interdomain routing system is described below with the component-based definition. 
\begin{itemize}
 \item The components set for the interdomain routing is $C = \{N1, N2, ..., N7\}$;
 \item The action space for each node is to deliver the package at hand to its neighboring nodes. Typical example is $A_{N1} = \{ toN2, $ $toN3 \}$;
 \item The only quality attribute this network needs to be concerned with is the time delivering the package to its destination as we assume there is no case of package loss. 
 Specifically, we consider the delivery time is proportional to the distance denoted by hops between nodes. 
 Its utility function is encoded using a formula that enables the quantification of the utility of a given state and defined as $U_{system} = 10 - \#hops$. 
 Usually, the longer time, the lower utility and the maximum utility system could achieve under normal operations for this network is 8 with two hops $\langle N1\ N2\ N5\rangle$.
\end{itemize}

Currently, $N2$ and $N4$ are analyzed to be potentially attacked based on the historical package delivery record, deliberately sending the package in the opposite direction, extending the delivery time.  
The game definition with the security attacks is summarized below. 
\begin{itemize}
\item The player set for the game is $C = \{N1, N2, ..., N7\}$. The set of affected components by the attack includes $N2$ and $N4$, i.e., $C_{att} = \{ N2,\ N4\}$;
\item The action set for all players, including malicious ones controlled by attacks, is delivering the package to its neighboring nodes;
\item The set of types for potential attacked component node includes ``normal'' and ``malicious'' (i.e., $\theta_{N2} \in \{normal,\ malicious\}$, $\theta_{N4} \in \{normal,\ malicious\}$);
\item The payoff for all the normal players is allocated by the system utility with the \emph{Shapley Value Method} (i.e., $U_{system} $ $\div\ |normal\ players|$, equally allocated in this case since all of the nodes in this network is not cut vertex with the same importance). 
For example. each node is awarded 8/7 if none of them is attacked. The utility for the ongoing attacks on two components is the utility loss from the system's best response without attack;
\item The probability distribution for both component $N2$ and $N4$ could be, e.g., 50\%/50\% split (i.e., $\rho_{N2, N4}(normal,$ $ malicious) = (0.5, 0.5)$. 
\end{itemize}


\begin{figure}
    \centering
    \includegraphics[width=5.5in]{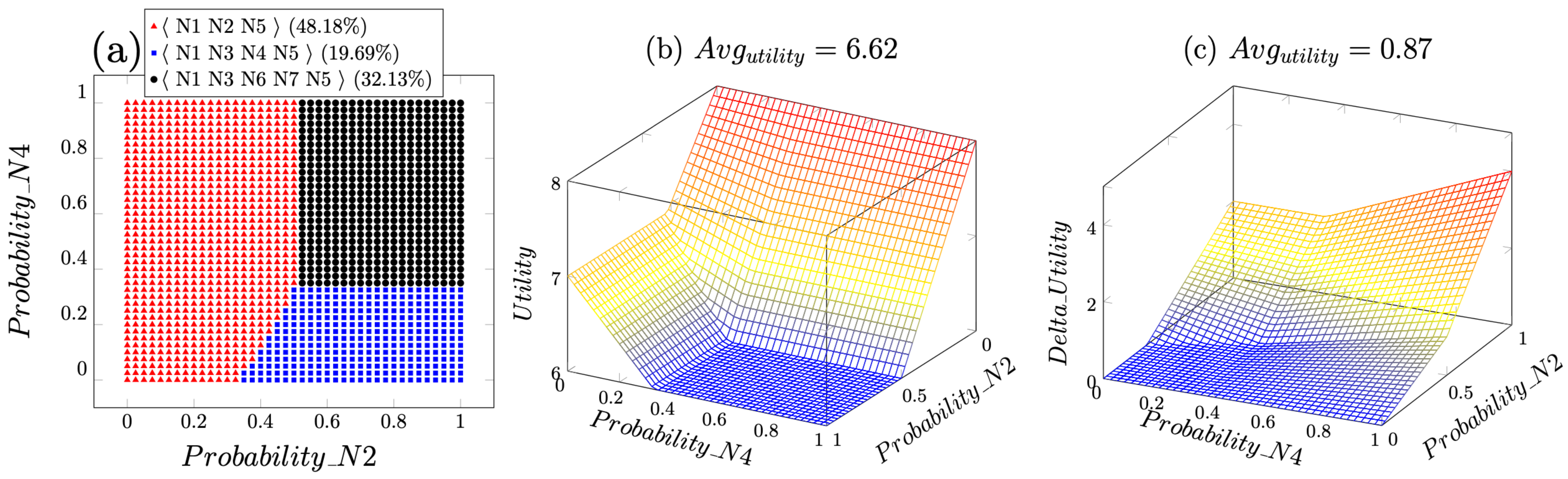}
    \caption{Results for interdomain route system: (a) expected route in equilibrium; (b) system utility with game theory approach; (c) delta between system utility from game theory approach and utility from greedy algorithm.}
    \label{routeresult}
\end{figure}

\subsubsection{Experiment Setup and Results}
\label{sec:eval2}
We demonstrate how our Bayesian game approach combined with the proposed dynamic programming algorithm can produce adaptation decisions about how to forward packages for each node in the routing example. 
Similar to the experiment results found on the Znn.com website, we statically analyzed a discretized region of the state space which represented different attack scenarios (i.e., malicious probability of $N2$ and $N4$). 
The entire experiment setup of the network structure is exactly shown in Fig. \ref{routingexample}(a). 
In addition, we also adopt a greedy algorithm for this routing application as the baseline, and compare the system utility between these two approaches to demonstrate the superiority of game theory under security attacks. 
The experiment for the whole state space with Bayesian approach takes less than one minute and the solution generation time for each state is negligible.

Fig.~\ref{routeresult} (a) presents the results of the policy selection (i.e., expected package sequence) over two dimensions that correspond to the malicious probability of $N2$ and $N4$, respectively. 
Red triangle points denote that the policy for $N1$ is $N2$, extending the range of $Probability\_N2$ to around [0, 0.50]. 
This is because when the chance of $N2$ coming under attack is less than 0.50, $N1$ should pass the package to $N2$, since $N2$ is in the shortest path to the destination; otherwise, $N1$ delivers the package to $N3$. 
Similarly, when the malicious probability of $N4$ is less than 0.35, the policy for $N3$ reaching equilibrium is to deliver the package to $N4$ (i.e., blue square points), since the benefits of a short delivery time outweigh the potential detriment. 
For the remaining situations denoted by the black circle points, $N1$ passes the package to $N3$, which in turn forwards it to $N6$.  

Fig.~\ref{routeresult} (b) describes the utility the system could obtain for the attacked components' equilibrium policies. As expected, when the $Probability\_N2$ is greater than 50\% and $Probability\_N4$ greater than 35\% (i.e., black circle points in Fig.~\ref{routeresult} (a)), the utility system can gain is 6 as there are 4 hops in the expected sequence $\langle N1\ N3\ N6\ N7\ $ $N5\rangle$). 
This plot also shows that the system utility increases progressively with decreasing probability of the compromised $N2$ and $N4$. 
When the $probability\_N2$ is 0, the expected utility increases to 8 (i.e., two hops in $\langle N1\ N2\ N5\rangle$). Similarly, the utility reaches 7 with $probability\_N4$ 0 (i.e., three hops in $\langle N1\ N3\ N4\ N5\rangle$). 

Furthermore, we adopt a baseline that generates policies for each node in a non-repeating fashion, passing the package to the adjacent node along the shortest path to the destination.
The aim is to compare the utility between two different approaches dealing with security attacks. 
For the network as shown in Fig.~\ref{routingexample}(a), the baseline firstly picks up the shortest path sequence $\langle N1\ N2\ N5\rangle$. 
If $N2$ is compromised and sends the package back, $N1$ redelivers it to $N3$ instead of $N2$ since the package is received from $N2$.  
The system utility for the greedy algorithm is the expected value, the weighted average of utility for paths in different attack situations. 
Fig.~\ref{routeresult} (c) shows the delta between the utility produced by our game theory method and the utility produced by the baseline.
During security attacks, we can see that the utility from the game theory approach is always higher than the greedy approach under security attacks. 
The delta is much more noticeable, especially in the situations where $N2$ and $N4$ are highly likely to be compromised (i.e., $Probability\_N2$ and $Probability\_N4$ close to 1). 
This is because game theory approaches can help the defenders to trade off the gains and losses due to perceived risks.





In summary, based on the preliminary results of our experiment, our game theory approach in the component level is applicable for self-adaptive applications. 
To adopt our approach, attacks information, such as various types with probabilities as well as its payoff, shall be provided from the \emph{Analyzer}, to construct a Bayesian game based on system architectural structures. 
The results have shown that game theory can enhance the performance of the system, especially when a potential attack is more likely to happen. 
In these situations, game theory approaches could help the defenders balance perceived risks by using underlying incentive mechanisms, and figure out the best response as the adaptation to be executed on the network using proven mathematics. 
Besides, our proposed dynamic programming algorithm is specific to this kind of application to optimize the game solving. 
Another potential application is the multi-agent finding (MAPF) problem where a spatial position in a path can be viewed as a node in the network~\cite{sukkerd2020tradeofffocused,amir2015multiagent}. 
Other optimization techniques might be adopted or customized for different applications with complicated game structures. 

\begin{figure}
    \centering
    \includegraphics[width=5in]{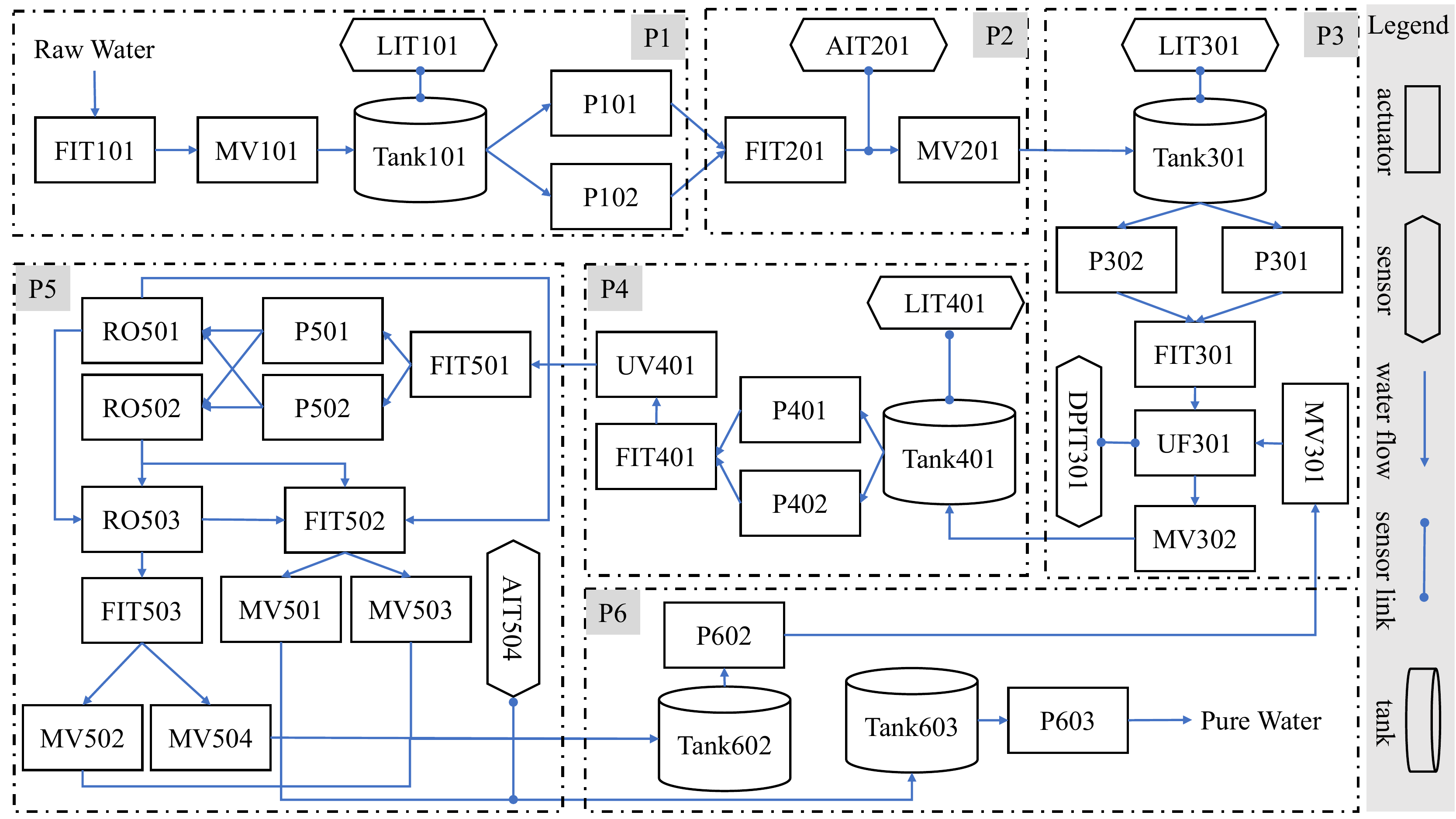}
    \caption{The physical components of SWaT system}
    \label{fig:runningexample}
\end{figure}

\subsection{A Case Study on SWaT System}
\label{sec:casestudy}
\subsubsection{SWaT System}
\label{sec:swat}
\emph{Secure Water Treatment} (SWaT) system is a room-scale testbed, which is a fully operational scaled-down version of modern water treatment plant found in cities, and it is designed and built for investigating responses to cyber attacks and experiments with novel secure approaches \cite{adepu2021distributed,kang2016model,siddhant2018design}.

Fig. \ref{fig:runningexample} illustrates the physical layer of the SWaT system, consisting of multiple sensors, actuators, and other physical devices. 
The control player of SWaT consists of multiple PLCs and the communication networks, which is the same as single tank example (shown in Fig. \ref{fig:singeltank}).
In the SWaT system, there are four kinds of sensors: LIT (tank level indicator/transmitter), AIT (analyzer indication transmitter), DPIT (differential pressure indicator/transmitter) and FIT (flow meter); five kinds of actuators: P (water or chemical dosing pump), MV (motorized valve), RO (reverse osmosis), UF (ultrafiltration), UV (ultraviolet dechlorinator); other physical devices: TANK (water or chemical dosing storage tank).
Some of these components are able to adjust their behaviors to the dynamics of the environment sensed by sensors at run-time.
For example, a Pump can be set to ON/OFF for allowing water to flow out of the corresponding tank or not when the water level in the tank is high/low; an MV can be set to Open/Close for allowing water to flow through the pipeline or not.

\begin{figure}
    \centering
    \includegraphics[width=4.5in]{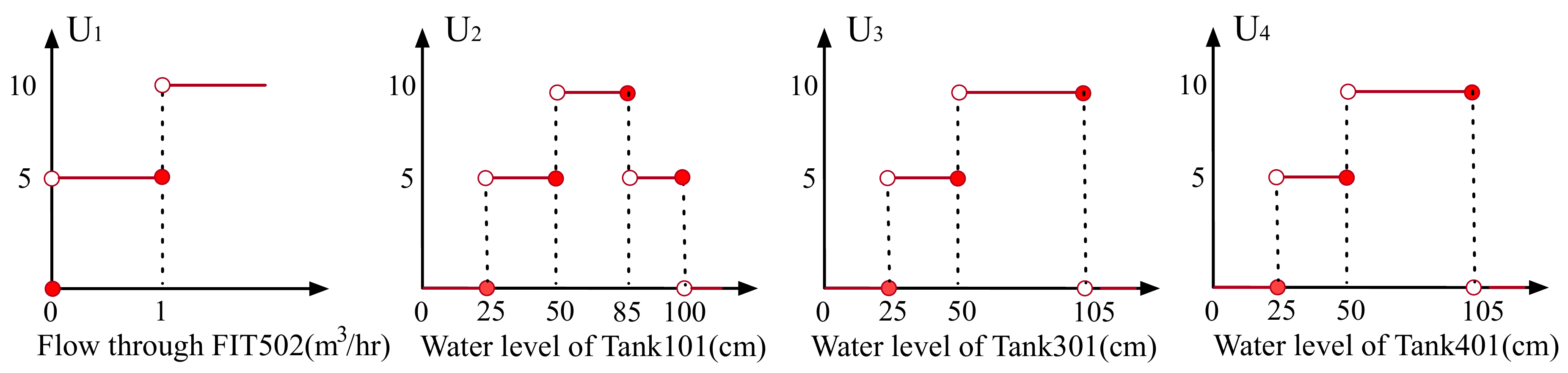}
    \caption{System-level utility functions of SWaT system}
    \label{fig:systemutility}
\end{figure}

The treatment process in SWaT system consists of six different stages (P1 through P6 in Fig.\ref{fig:runningexample}), in which the SWaT system employs various sensors and actuators that monitor and manipulate the state of physical equipment, such as water or chemical dosing tanks.
Through the ``cooperation'' of all the components in each stage, the SWaT system takes as input raw water and undergoes various chemical treatments to output filtered water in the last stage.
However, the ideal cooperation may be broken under security attacks. 
One of the potential attacks is to control or mislead a component by compromising the communication link between this component and its managing PLC.
For example, an attacker invades the link between LIT101 and its managing PLC, and sets LIT101's value to $200mm$ (but the true value is already $800mm$). 
$200mm$ is lower than the predefined value (i.e., standard\_level in Listing 1); hence, P101 turns OFF and does not allow water to flow out of Tank101 which results in the tank overflow.
The system-level utility designed by experts for SWaT system is: $U_0=\sum_i^{4}U_i$, where $U_i$ is shown in Fig.\ref{fig:systemutility}.
Here, $U_i$ denotes $i$th term utility rather than payoff for component $i$.
In Fig. \ref{fig:systemutility}, the x-axis represents the status read by corresponding sensors, and the y-axis represents the utility value.

\subsubsection{Self-Adaptation for SWaT}
Adaptation behaviors are built on the equilibrium responding to unexpected attacks and are achieved by elaborating the proposed self-adaptive framework (shown in Fig. \ref{fig:overviewfigure}).

Concretely, \emph{Knowledge Base} stores the necessary information for the sake of self-adaptation, including (1)-(5) mentioned in Section \ref{sec:overview-2}, where (4) and (5) are stored in the form of knowledge case defined in Definition \ref{def:kcase}, i.e., probabilities of components being successfully compromised and the joint action derived from the corresponding equilibrium policy.
Quality attributes are defined in terms of the overflow or underflow state of a tank and water throughput of SWaT.
\emph{Monitoring Activity} gathers and synthesizes the on-going attack information through sensors and saves information in the Knowledge Base. 
For instance, SWaT will monitor typical operational data, including LIT101, FIT101, MV101, P101, etc. 
\emph{Analysis Activity} performs analysis and further checks whether certain components are attacked with probabilities; potential deviated malicious actions are identified; the rewards for the attack are estimated, based on the knowledge about component vulnerabilities and system objectives. 
The deep learning-based predictor designed in Section \ref{sec:att} is adopted to identify the anomalies with probability prediction caused by security attacks. 

\emph{Planning Activity} generates adaptation policies by solving the Bayesian game, constructed in an automatic way from system architecture model with the input of potential attacks and system objectives. 
The high-level view of the whole planning process is shown in Fig.\ref{fig:construct}.
Based on the description of the system architecture and the prediction generated by \emph{analysis}, process (1) retains the core information by tailoring away unnecessary components such as FIT sensors and storing the anomaly information. 
Process (2) uses a topological sort algorithm \cite{cormen2009introduction} to generate the order of play that determines which component takes action first and which one takes action later for Bayesian game construction.
Process (3) uses a SWaT simulator to calculate utilities for all possible joint actions, and assign payoff to each component based on the Shapley value method.
Process (4) takes as input the order of play and payoff, and outputs a Bayesian game in XML format. 
Process (5) finds a Nash equilibrium as the adaptation to potential attacks by putting the game constructed into a game solver. 
The game solver we adopt is \emph{Gambit} \cite{gambit}.
Adaptations from equilibrium are enacted during  \emph{Executing Activity} through actuators. 

\subsubsection{Experiment Setup and Results}
The comprehensive case study on the SWaT system demonstrates that the self-adaptation policy changes with the compromised probability so that the system can effectively respond to the attacks and minimizes the potential harm to the system.

\begin{figure}
    \centering
    \includegraphics[width=5in]{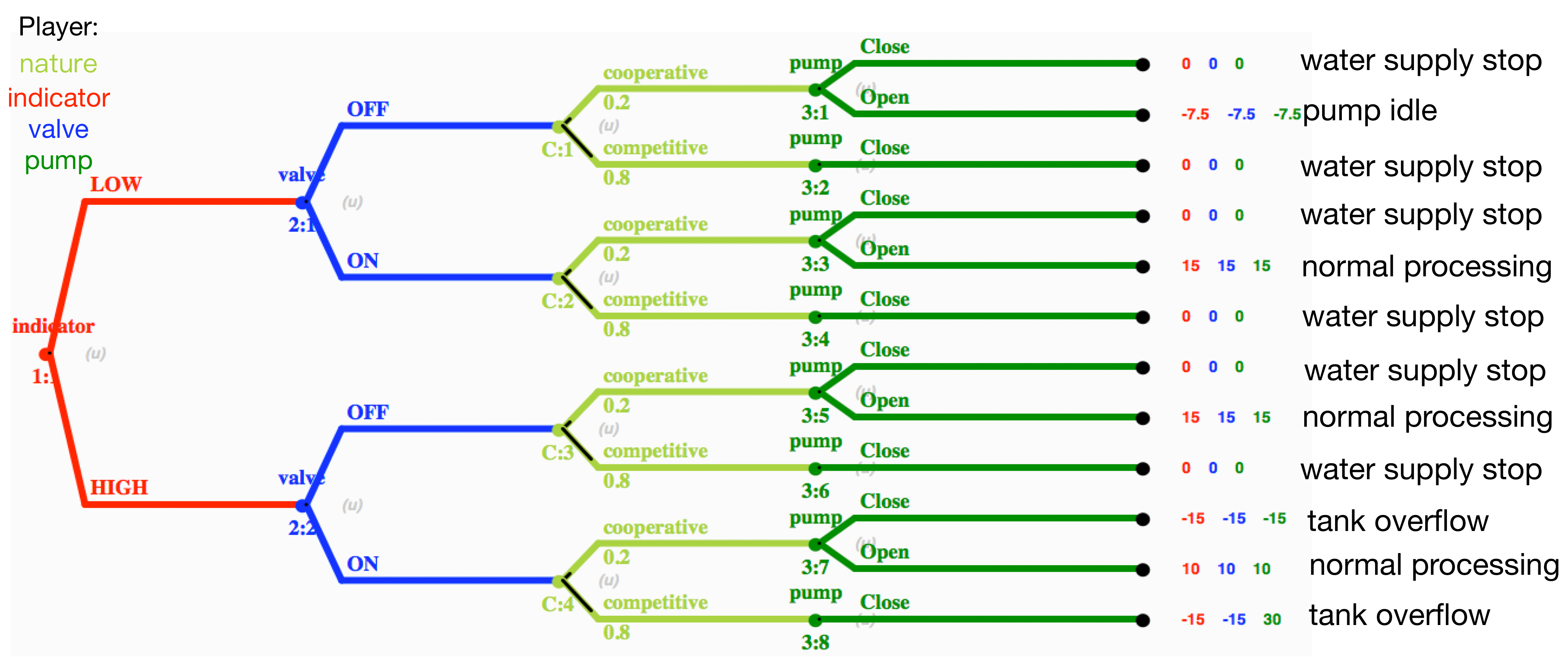}
    \caption{The game tree for the single tank system under attack A1. In the figure, above each node is the name of the component, and below each node is the name of information set (the same in Fig.\ref{fig:attack2}).}
    \label{fig:attack1}
\end{figure}

\textbf{Single Tank Experiment.}
In the single tank experiment, three components are modeled, i.e., motorized valve, outlet pump, and water level indicator.
The action sets of them are $\{\text{OFF},\text{ON}\}$, $\{\text{Close},\text{Open}\}$, and $\{\text{LOW},\text{HIGH}\}$, respectively.
Note, action ``LOW'' and ``HIGH'' are the signals that the water level indicator sends to others.
If the indicator is not compromised, once it detects that the water level in the tank is lower/higher than $500mm$, it sends the signal ``LOW''/``HIGH'' to others.
Otherwise, once it detects that the water level in the tank is lower/higher than $500mm$, it sends the signal ``HIGH''/``LOW'' to others.
To simplify the utility function, the the system level utility in the running example is set to: (1) tank overflow, -30; (2) water supply stop, 0; (3) pump idle, -15; (4) normal processing, 30. 
And the Shapley value-based approach designed in section \ref{sec:component-payoff} is used to deduce the payoff for each component from the system level utility.

\begin{figure}
    \centering
    \includegraphics[width=6in]{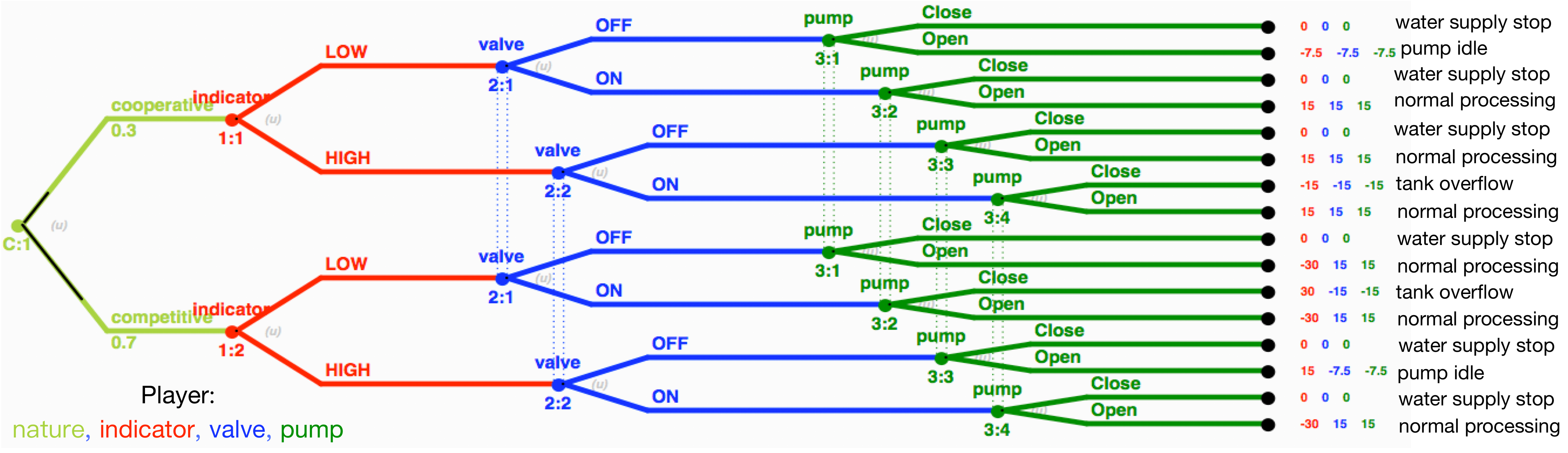}
    \caption{The game tree for the single tank system under attack A2. The two nodes connected by a dotted line are in the same information set.}
    \label{fig:attack2}
\end{figure}

Two attack scenarios are considered, i.e., keep the outlet pump always close (Attack A1 shown in Table \ref{attakctype}) and falsify the status of the water level  (attack A2 shown in Table \ref{attakctype}).
Fig.\ref{fig:attack1} illustrates the game tree generated by our approach for the single tank system under attack A1.
The game plays as follows: first, the indicator detects the water level and takes action LOW or HIGH; second, the valve takes action OFF or ON; third, the nature stochastically determines whether the attack on the pump is successful or not, i.e., choosing the type of the pump (competitive or cooperative); fourth, the pump takes action, if the pump is cooperative, it takes Close or Open based on its payoff; otherwise, it takes Close.
Fig.\ref{fig:attack2} illustrates the game tree generated for the single tank system under attack A2.
The game plays as follows: first, the nature stochastically determines whether the attack on the indicator is successful or not, i.e., choosing the type of the indicator (competitive or cooperative), and different types of indicators have different payoffs; second, the indicator takes action LOW or HIGH; third, the valve takes action OFF or ON, and it do not know the type of the indicator, i.e., unknown to the truth of the water-level signal; fourth, the pump takes action, and it does not know the truth of the water-level signal either.

\begin{figure}
    \centering
    \includegraphics[width=4in]{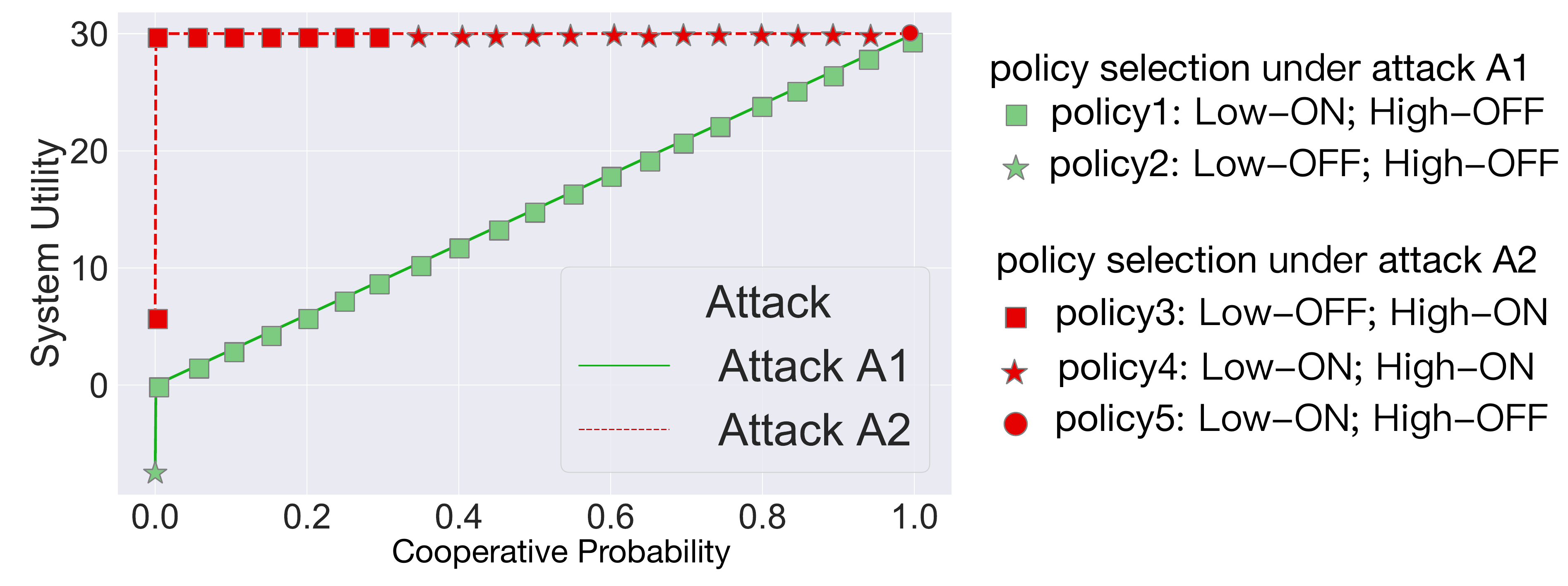}
    \caption{Experimental results: the system utility of single tank system under two typical attack scenarios (attack A1 and attack A2); the equilibrium policies for motorized valve under A1 and A2, respectively.}
    \label{fig:attackresults}
\end{figure}

Fig. \ref{fig:attackresults} shows the system utility and the policy selection for the motorized valve under different cooperative probabilities of the pump and indicator.
For the green line, the x-axis represents the cooperative probability of the pump.
For the red line, the x-axis represents the cooperative probability of the water level indicator.
The policy ``Low-ON; High-OFF'' means that if the water level is low, valve turns on, and if the water level is high, valve turns off. 
As expected, the system utility increases with the increasing cooperative probability of pump/indicator.
If all components are normal, the optimal policy of the valve is ``Low-ON; High-OFF'';
if the pump is successfully compromised, the optimal policy is ``Low-OFF; High-OFF'', which stops water supply to avoid tank overflow; if the indicator is successfully compromised, the optimal policy is ``Low-OFF; High-ON'', which means that the valve knows that the water level signal is the opposite of what the level really is.
And these policies are consistent with expert decision-making.
Notice that the pure Nash equilibrium is usually not unique. 
For example, in the game as shown in Fig.\ref{fig:attack2}, when the cooperative probability of the indicator is set to $0.2$, there are $31$ pure Nash equilibria. 
Fig.\ref{fig:attackresults} shows one of the results. 
The multiple equilibria problem will be discussed in Section \ref{sec:con}.

\textbf{System-Wide Experiment.}
For constructing a Bayesian game of the whole SWaT system, the five steps shown in Fig. \ref{fig:construct} are conducted as follows:
(1) Only those actuators with run-time actions such as motorized valve and pump will be modeled as players; a pair of replaceable pumps, such as P101-P102, P501-P502, will be modeled as one player for simplicity.
Here, the components modeled as players are MV101, P101, MV201, P302, MV302, P402, P501, and MV501; a modeled components graph is constructed (as shown process (1) in Fig.\ref{fig:construct}).
(2) The order of play with \emph{nature} player is generated.
For example, if the potentially compromised components are MV101 and P402, the order of play is [nature, MV101, P101, MV201, P302, MV302, nature, P402, P501, MV501];
if the potentially compromised components are MV302, the order of play is [MV101, P101, MV201, P302, nature, MV302, P402, P501, MV501].
(3) A SWaT simulator and Shapley value method are used to generate the payoff for each component under each possible joint action.
(4) Algorithm \ref{algorithm_add} and Algorithm \ref{algorithm_dlr} are used to generate the Bayesian game in an XML formation that Gambit can solve.
(5) Gambit is used to solve the game.
Moreover, we use function $U_0$ mentioned in Section \ref{sec:swat} to compute system level utility.

The bottom left of Fig.\ref{fig:construct} illustrates an example of the Bayesian game: there are two components, an abnormal component MV101 (in red) and a normal one P101 (in blue); the probability of MV101 being compromised is $0.2$; the two nodes connected by a dotted line are in the same information set; the black nodes are terminal nodes, and the two numbers behind the terminal node represent the payoff for MV101 and P101; 
all the dark edges constitute the equilibrium path, i.e., (1) if MV101 is not compromised, MV101 will take Open and P101 will take ON, and they get $20$ and $20$ payoff, respectively; (2) otherwise, MV101 will take Close and P101 will take OFF, and they get $-17$ and $17$ payoff, respectively.

\begin{figure}
    \centering
    \includegraphics[width=6.55in]{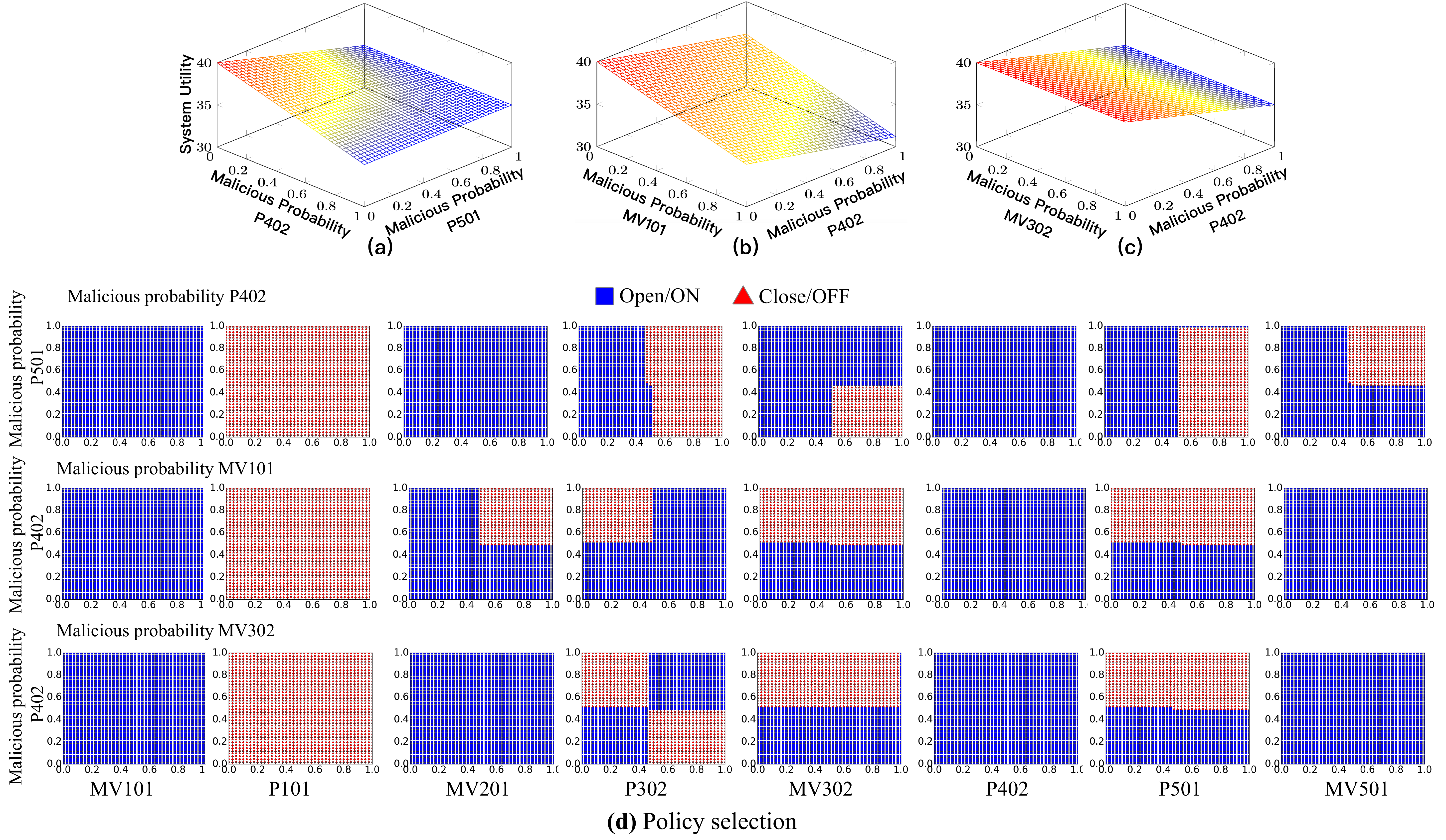}
    \caption{Experimental results: the system utility under three typical attack scenarios(shown in (a),(b),(c)), and the equilibrium policies for eight components under the three typical attack scenarios (shown in (d)).}
    \label{fig:rq1}
\end{figure}

Fig. \ref{fig:rq1} shows the equilibrium results of the Bayesian game for SWaT system. 
Here, we statically analyzed a discrete region of the state space which represented different attack scenarios, i.e., malicious (competitive) probability of two components.

We try $C^2_8=28$ pairs of abnormal components to investigate the impact of different attacks on the system utility (which is equal to the sum of all normal components' payoffs).
Fig.\ref{fig:rq1} (a)-(c) present three typical cases of the system utility over two dimensions that correspond to the malicious probability of [P402, P501], [MV101, P402], and [MV302, P402], respectively.
In Fig.\ref{fig:rq1}(a), when malicious probability of [P402,P501] are [1.0,0.0], [0.0,1.0], and [1.0,1.0], the system utilities are always 35.00. This means that only-P402 attack, only-P501 attack, and both-P402-P501 attack have the same effect on the system utility.
In Fig.\ref{fig:rq1}(b), when malicious probability of [P402,MV101] are [1.0,0.0], [0.0,1.0], and [1.0,1.0], the system utilities are 35.00, 36.17, and 31.17, respectively. This means that attacking both MV101 and P402 at the same time would have a worse effect on the system than single-component attacks.
In Fig.\ref{fig:rq1}(c), when malicious probability of [MV302,P402] are [1.0,0.0], [0.0,1.0], and [1.0,1.0], the system utilities are 40.00, 35.00, and 35.00, respectively. This means that attacking MV302 has no effect on the system.
Based on these results, we further suggest that some components of the SWaT system (such as P402, MV101) need to be more strictly protected, while some components can be paid less attention (such as MV302).

\begin{figure}
    \centering
    \includegraphics[width=6in]{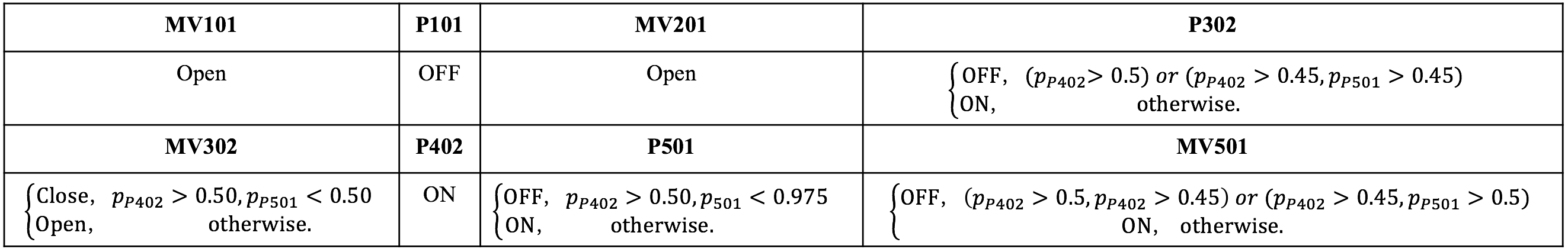}
    \caption{Rules for policy selection when P402 and P501 are under attack.}
    \label{fig:rule-for-policy}
\end{figure}

Fig.\ref{fig:rq1}(d) demonstrates the results of the policy selection of the eight components under the above three attack scenarios.
The eight figures in the first/second/third row in Fig.\ref{fig:rq1}(d) correspond to the attack scenario of Fig.\ref{fig:rq1}(a)/(b)/(c).
There are two typical patterns in policy selection for each component:
(1) ``Static policy'', i.e., the policy selections of the components remain the same no matter how the malicious probabilities of the two compromised components change. MV101, P101, and P402 hold the static policies;
(2) ``Dynamic policy'', i.e., the policy selections of the components are sensitive to the malicious probabilities of compromised components. MV201, P3022, P501, and MV501 hold the dynamic policies.
Notice that although attacking MV302 has no effect on the system, it still changes its policy with the malicious probabilities of compromised components to maximize the system utility.
Further, based on the patterns, we can design a set of rules to determine the policy selection rather than solving the Bayesian game.
For instance, under the attack scenario shown in Fig.\ref{fig:rq1}(a), i.e., P402 and P501 are under attack, the rules of policy selection for each component can be designed as shown in Fig.\ref{fig:rule-for-policy}.

In this case study, although one limitation of our evaluation is that we did not directly compare our analytical results with other approaches in terms of performance and effectiveness for SWaT system, Section \ref{sec:e1} and \ref{sec:e2} performed comparison on website application and routing application with the baseline of stochastic model checking and greedy solution approaches, where component-based game approaches indeed further improve the overall system utility.

\textbf{Performance overhead.}
The time-consuming is twofold.
One is the offline time which includes the time of generating the order of play (0.01 sec), utility generating in the simulator (11.53 sec), generating the Bayesian game (0.05 sec), solving the game (99.33 sec),
and training the compromised probability predictor (625 sec). 
To cope with different cases for potential attacks, we can sample the games on a discrete region of state space and compute their equilibria for each state in advance.
The other is online adaptation time which includes the time of regenerating the malicious probability (1.89 sec), loading the policy from the knowledge base (0.01 sec).
In addition, the time-consuming is intuitively proportional to the number of the components. From 5 to 8 components, the time of solving the corresponding game is 0.75 sec, 3.57 sec, 15.85 sec, and 99.33 sec. 
To improve the scalability of our approach, please refer to the dynamic programming algorithm proposed in Appendix \ref{sec:dp} by decomposing a game into sub-games.

Moreover, as mentioned above, there are some typical patterns in the policy selections. Based on these patterns, a set of rules could be designed to determine the policy selection, which can significantly reduce computing and storage costs. 
Besides manually designing the policy-selection rules, machine learning algorithms, such as decision-making tree, support vector machine (SVM), and deep neural networks, can also be used to generate these rules.

\section{Conclusion}
\label{sec:con}
In this paper, we propose an approach for securing software-intensive systems using a rigorous game-theoretical framework by extending a self-adaptation framework with component-level Bayesian game. 
The algorithm has been designed to generate the optimal defense policy from the equilibrium by automatically solving the game when detecting potential attacks. 
We have evaluated our approach on three systems, i.e., Znn.com system, interdomain routing system and SWaT system, and the evaluation shows the applicability and the effectiveness of our approach.
And its applicability can be further enhanced by applying it to more practical scenarios in other domains. 

It is noticed that we simplify the modeling of the threats for easing the representation, for example, restricting the number of component types under attacks and assuming the attackers with fully competitive behaviors.
And the constructed game is \emph{one-shot-play}, that is, each player only takes action once in the game. 
In real-world security landscape, it could be more complex.
For example, the security attacks with highly motivated and capable adversaries willing to devote significant time and continuous attack to facilitate their malicious goals \cite{DBLP:conf/memocode/KinneerWFGG19}. The one-shot-play game may fail to depict such advanced persistent threats (APTs).
Accordingly, in the future, we are planning to extend our approach to incomplete-information \emph{Markov game} that might be able to model the multi-stage interaction between a set of players with uncertainties.

Moreover, in this work, we adopt pure equilibrium as the adaptation response. 
However, in practice, there will likely be multiple equilibria and no guarantee of uniqueness.
While this is an area for future work, one possible way to overcome this is to adopt other solution concepts in game theory (such as \emph{perfect Bayesian equilibrium}, \emph{trembling-hand perfect equilibrium}, or \emph{strong Stackelberg equilibrium}).
Another limitation, and also a topic for future work, is the mixed equilibrium which might be another solution for game theory. 
Its interpretation on system behaviors could be various and allows generation of different types of defense policies for the system, which can be explored for different applications.

\newpage

\appendix
\section{Appendix: Dynamic Programming Algorithm for Routing Games}
\label{sec:dp}
In practice, a network might be complex and each node could have hundreds of neighboring nodes. It is impractical to directly build a Bayesian game tree, in the component level with a large number of players (each with a massive action set), and solve such a network in a reasonable time. 
To deal with the complexity of network nature, we propose an algorithm inspired by dynamic programming to effectively solve the generated Bayesian game for this class of routing problems.

The algorithm relies on the following functions and variables:
\begin{itemize}
    \item $dis(i,d)$: labeling the distance between node $i$ and the destination node denoted by the number of hops in the shortest path, e.g., $dis(N1,N5) = 2$.
    \item $disValue$: indicating the distance within which some nodes are to be found. 
    \item $uncertain(i)$: judging whether node $i$ is uncertain and attacked leading to two types, e.g., $uncertain(N4) = true$, $uncertain(N3) = false$. 
    \item $adj(i)$: returning the set of nodes adjacent to node $i$, e.g., $adj(N3) = \{N1,\ N4,\ N6\}$.  
    \item $buildGame(i)$: constructing a sub-game starting from $i$, usually adjacent to an uncertain node.
    \item $solve(gambitTree)$: solving the sub-game and figuring out its policy with equilibrium.
    \item $addNode(e, S)$: adding a node $e$ to set $S$.
    \item $removeNode(e, S)$: deleting the node $e$ from set $S$. 
    \item $subG$: the set of nodes which have been found their best reactive actions.
    \item $todoS$: the set of nodes which should be solved with their best policy but not yet due to some of their unprocessed neighbouring nodes. 
\end{itemize}

The Algorithm~\ref{algorithm1} for routing game 
has as input a routing network $N$ -- consisting of a starting point $s$ of package delivery and a destination point $d$. 
To carry out dynamic programming, the algorithm uses a set $subG$ to store the set of nodes which have been processed with their best reactive policy.
$subG$ is initialized as an empty set (line 1) and added with node $d$ (line 2) since $d$ does not need the policy to transmit the package. The algorithm starts by iterating all the nodes in the distance $disValue$ (line 5), initialized by 1 (line 3). For example, $N2$, $N4$ and $N7$ are qualified in the first iteration. Each node is checked whether it is potentially attacked (i.e., $uncertain(n)$ in line 6). For those uncertain nodes (e.g., $N2$ and $N4$), they might affect the policy of their prior nodes (line 7) (e.g., $N1$ and $N3$), which shall be added to $todoS$ (line 8), to be processed to update their policy due to its neighboring uncertainty. A typical example is that node $N3$ might trade off the delivery between $N4$ and $N6$ even though $N4$ is in the shortest path from $N3$ to $N5$, however, could deliberately send the package back controlled by the attack. If the node is not in $todoS$ to be updated (line 11), it is directly added to the $subG$ (line 12) as the best policy for such benign node is passing the package down to its adjacent node along the shortest path. In this routing scenario, $N2$, $N4$ and $N7$ are added to $subG$ as their policies in equilibrium with normal type are easily determined. 

After iterating all the nodes in $disValue$ 1, each node in $todoS$ (line 15) is checked whether it satisfies the condition (line 16) where all its neighboring nodes (i.e., $i\in adj(n)$ ) closer to destination (i.e., $dis(i,d)==dis(n)-1$) have been solved with their best policies (i.e., in $subG$), to build a sub-game. As shown in the example, though both $N1$ and $N3$ are prior to an uncertain node, their policy update is postponed as $N6$ is not in $subG$ yet, which affects the sub-game generation for $N3$, in turn delaying the sub-game construction for $N1$. 

\begin{algorithm}[htb]
\algsetup{linenosize=\small}
\caption{Dynamic Programming Algorithm to Solve Routing Game.}
\label{algorithm1}  
\begin{algorithmic}[1]
\STATE $subG \Leftarrow \emptyset$
\STATE $addNode(d, setG)$
\STATE $disValue \Leftarrow 1$
\REPEAT  
\FORALL {$n \in N\ \ and\ \ dis(n,d)==disValue$} 
 \IF{$uncertain(n) == true$}
   \FORALL{$n_p \in adj(n)\ \ and \ \ dis(n_p,d)==disValue+1$}
   \STATE   $addNode(n_p, todoS)$
   \ENDFOR
  \ENDIF
 \IF {$n\notin todoS$} 
 \STATE $addNode(n, subG)$
 \ENDIF
\ENDFOR
\FORALL{$n\in todoS$}
\IF{$\forall i\in adj(n)\ and\ dis(i,d)==dis(n)-1\ and\ i\in sutG$}
\STATE $gambitTree \Leftarrow buildGame(n)$
\STATE $equilibria \Leftarrow solve(gamebitTree)$
\STATE $removeNode(n, todoS)$
\STATE $addNode(n, subG)$
\ENDIF
\ENDFOR
\STATE $disValue \Leftarrow disValue + 1$
\UNTIL{$s\in subG$} 

\end{algorithmic} 
\end{algorithm}

An exemplified subgame construction (line 17) starting from $N3$ is illustrated in Fig.\ref{routingexample}(b) when all conditions are satisfied. The stochastic behavior of those potentially compromised nodes can be modeled by introducing a nature (or chance player), who moves according to the probability distribution (e.g., 50\%/50\% split), randomly determining whether attacks on $N2$ and $N4$ are successful. Then, $N3$ can choose an action passing to the one from the set of its adjacent nodes, i.e., $N6$ or $N4$. Here, $N3$ is a normal node aware of that the package is transmitted from $N1$ and it is not necessary to consider a rollback to $N1$. The game is ended after $N3$'s action as we can prune the following branches: 1) to $N6$, the remaining route sequence is $N7$ and $N5$ by default as their best policy have been solved (i.e., $N6$  delivers the package to $N7$, which in turn forwards to $N5$);
2) to $N4$, with $N4$ forwarding to $N5$ if it is normal while backing to $N3$ in malicious type. When the game terminates, each player gets a unique payoff following different branches. As shown in the left most rectangle
all the players (including $N2$ and $N4$ as they are benign collaborating nodes) equally share the system utility value 6 with 3 hops from $N3$ to $N5$ plus the shortest path from $N1$ to $N3$. However, on the rightmost branch, only five players ruling out $N2$ and $N4$ is allocated with the system utility 4. The system utility is resulting from 6 hops if $N3$ decides to deliver the package to $N4$ as the nature problematically chooses the malicious type for $N4$, which sends the package back to $N3$ to maximize the attack's utility. Once $N3$ receives the package from $N4$, it redelivers the package to $N6$ because $N3$ as a good player does not repeatedly send it back. To this end, $N2$ and $N4$ are uniformly allocated the delta (i.e., 4) between the utility system obtained (i.e., 4) and the maximum utility system could obtain (i.e., 8) as the payoff. The payoff of the remaining branches can also be calculated accordingly.

After that, a pure Nash equilibrium is generated by solving this sub-game (line 18)  with Gambit software tools~\cite{gambit}, and the best policy for the node is updated according to the  equilibrium. By solving the sub-game for $N3$, the policy for $N3$ in the equilibrium is to deliver the package to $N6$, as the potential detriment on delayed delivery time to $N4$ due to attacks is greater than its comparative advantage of the shortest path.  Thus, this node with the solved policy is removed from $todoS$ (line 19) and absorbed in $subG$ (line 21).  Once all the nodes in the distance of $disValue$ from the destination have been iterated and all the nodes in $todoS$ satisfying conditions are computed for their best policy, the algorithm increment the value of $disValue$ one unit (line 23) and continue, until the starting point $s$ is in the set $subG$ (line 24). 

\end{document}